\pdfoutput=1
\documentclass[12pt,a4paper]{article}
\usepackage[left=1in,right=1in,top=1.5in,bottom=1.5in]{geometry}
\usepackage{mathrsfs}
\usepackage{amssymb}
\usepackage{amsmath}
\usepackage{ascmac}
\usepackage{amsthm}
\usepackage{graphicx}
\usepackage{CJKutf8}
\usepackage{booktabs}
\usepackage{enumerate}
\usepackage{setspace}
\usepackage{multirow}
\usepackage{natbib}
\usepackage{color}
\usepackage{xcolor}
\usepackage{graphbox}
\usepackage[normalem]{ulem}
\usepackage{url}
\usepackage{bm}
\usepackage{comment}
\usepackage[colorlinks,linkcolor=blue,citecolor=blue,urlcolor=blue]{hyperref} 
\usepackage{times}
\usepackage{algpseudocode}
\usepackage{algorithm}
\usepackage{titlesec}
\titleformat*{\section}{\large\bfseries}
\titleformat*{\subsection}{\it}
\titleformat*{\subsubsection}{\it}

\newtheorem{prp}{Proposition}

\def\cellbox#1#2{%
\parbox[c]{#1}{\baselineskip=0.7\baselineskip\rule{0pt}{1.1em}\noindent
#2%
\rule[-7pt]{0pt}{1em}}%
}
\def\tc#1{\multicolumn{1}{c}{#1}}

\makeatletter
\renewcommand\paragraph{\@startsection{paragraph}{4}{\z@}%
                                    {0.25ex \@plus0ex \@minus.25ex}%
                                    {-1em}%
                                    {\normalfont\normalsize\bfseries}}
\makeatother

\title{{\bf Spatially Dependent Indian Buffet Processes\footnote{This version: \today}
}}
\author{
Shonosuke Sugasawa$^{1}$ and Daichi Mochihashi$^2$
}
\date{}

\begin{document}

\doublespacing

\maketitle

\vspace{-3em}
\begin{center}
$^1$Faculty of Economics, Keio University\\
$^2$The Institute of Statistical Mathematics
\end{center}

\vspace{0.2cm}
\begin{center}
{\large\bf Abstract}
\end{center}

We develop a new stochastic process called spatially dependent Indian buffet processes (sIBP) for binary feature matrices of unbounded columns with spatial correlations between subjects, and propose general spatial factor models for various multivariate response variables. We introduce spatial dependency through the stick-breaking representation of the original Indian buffet process (IBP) \citep{griffiths05ibp,griffiths2011indian} and latent Gaussian process for the logit-transformed breaking proportions to capture underlying spatial correlation. We show that sIBP retains the sparsity and finite-feature behavior of the original IBP, while its joint feature allocation probabilities are affected by spatial correlation. Using binomial expansion and P\'olya-gamma data augmentation, we provide an efficient Gibbs sampler for posterior computation. The usefulness of our sIBP is demonstrated through simulation studies and two applications for large-dimensional multinomial data of areal dialects and geographical distribution of multiple tree species.


\newpage

%--------------------------------------------%
%              Introduction                  %
%--------------------------------------------%
\section{Introduction}

\begin{figure}[t]
 {\small
 \begin{tabular}{cc}
  \includegraphics[align=c,scale=0.2]{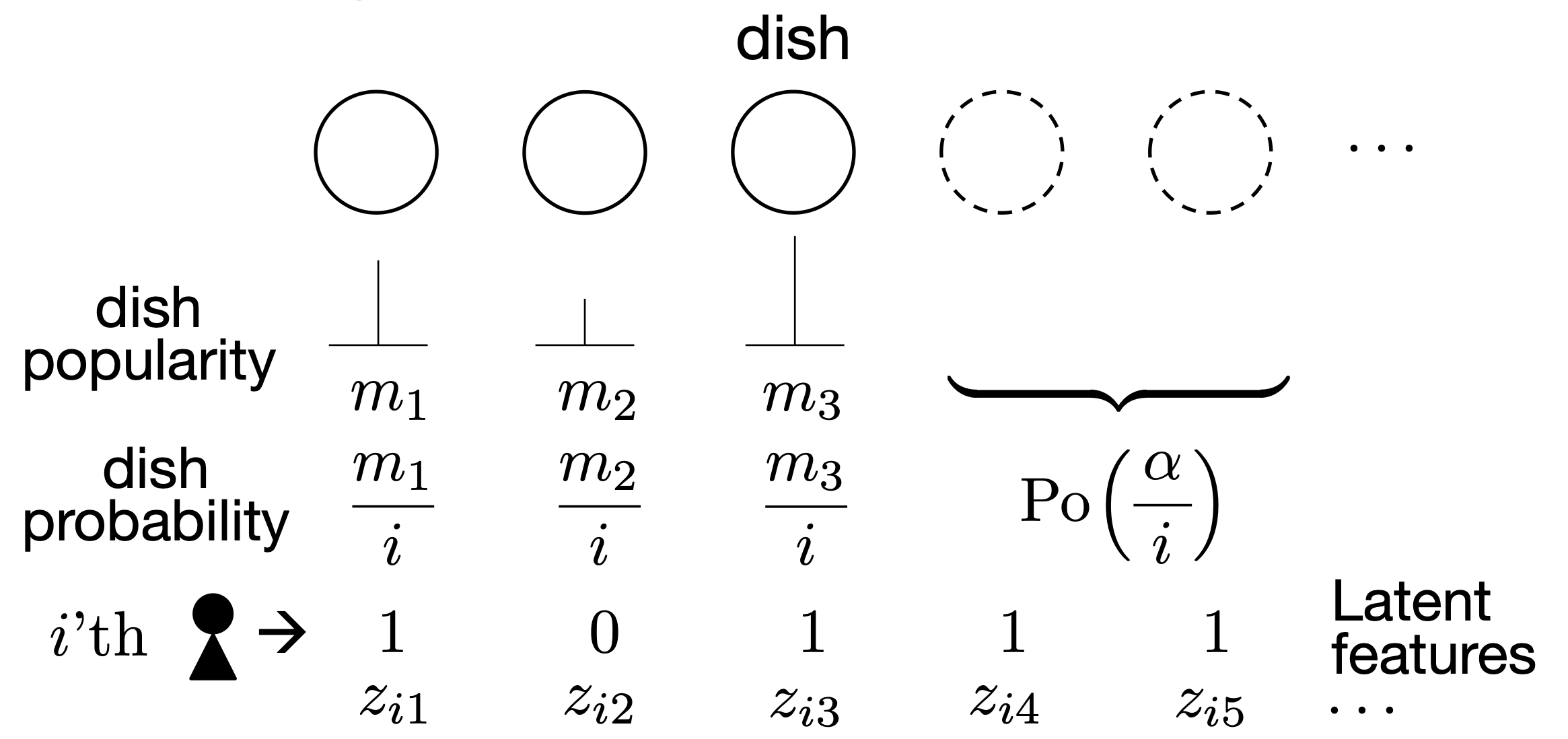}
  &
  \includegraphics[align=c,scale=0.5]{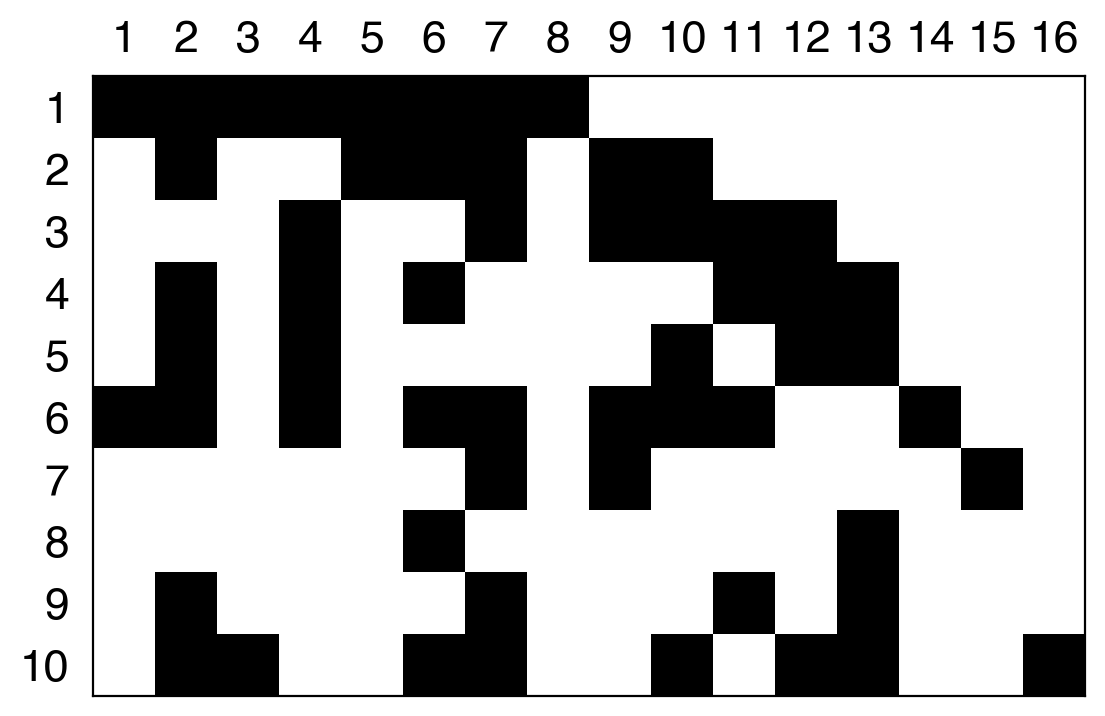}
  \\
  (a) Chinese restaurant process to generate latent features.
  &
  (b) Sampled latent feature matrix $Z$.\rule{0pt}{1.5em}
 \end{tabular}
 }
 \vspace{-0.3em}
 \caption{Indian buffet process (\cite{griffiths2011indian}).}
 \label{figure:ibp}
\end{figure}

The Indian buffet process (IBP) \citep{griffiths05ibp} is a powerful model for representing infinite binary matrices of data and features as shown in Figure~\ref{figure:ibp}(b),
particularly useful for factor models of high-dimensional data.
One of the key advantages of the IBP is its ability to automatically determine the number of latent factors, making it a convenient tool for dimensionality reduction and feature extraction. 
In addition, factor analysis using the IBP can also yield clustering according to the pattern of latent features.
Unlike traditional methods like $K$-means clustering, the binary representations proposed by IBP allow for capturing similarities between clusters and representing numerous clusters with only a small number of factors. 
See \cite{griffiths2011indian} for an introduction and review of IBP. 

One of the potential limitations of the vanilla IBP is, however, that it does not incorporate spatial information
associated with each data point.
This omission may lead to overlooking spatial heterogeneity and correlations, potentially resulting in inadequate factor representations. For example, when analyzing dialect data collected from various regions (a high-dimensional categorical dataset) as used in Section~5.1, it is natural to assume that geographically close areas would share similar linguistic structures. 
However, standard IBP models cannot capture such spatial information. Therefore, there is a need to develop a generalized version of IBP that can account for geographical relationships and spatial dependencies for successfully applying the IBP to multivariate spatial data.

To address this issue, we introduce a spatially dependent Indian buffet processes (sIBP) using stick-breaking representations, which can effectively incorporate geographical information into the modeling of multivariate binary factors. 
This approach enhances the flexibility and accuracy of the model in capturing spatial dependencies and automatically determines the optimal number of latent factors needed for a given dataset. 
To complement this modeling framework, we develop a highly efficient Markov Chain Monte Carlo (MCMC) computation algorithm. 
This algorithm leverages a novel data augmentation strategy that combines the binomial expansion and well-known P\'olya-gamma augmentation \citep{polson2013bayesian}, which enables to carry out the posterior computation by a Gibbs sampler.

There have been several attempts to extend the standard IBP models. 
The dependent Indian buffet process \citep{williamson2010dependent} introduces a stochastic process for a binary matrix $Z(t)$ indexed by $t$, which differs from our proposal as our method introduces spatial correlation within the elements of a binary matrix $Z$.
\cite{stolf2024allowing} proposed the multivariate probit Indian buffet process that accounts for correlations among different features but assumes independence among subjects, unlike our proposal.
\cite{warr2022attraction} and \cite{gershman2014distance} introduce the distance-based Indian buffet process, which incorporates distance information using the Chinese restaurant representation for IBP. 
However, it defines a multivariate model only for observed locations, and construction is not process-based, making the prediction of unobserved locations unjustifiable. 
There have been other various works on generalizations of the classical IBP model \citep[e.g.][]{broderick2013feature,di2020non,camerlenghi2024scaled}, but to the best of our knowledge, this work is the first to propose a process-based model that generalizes the standard IBP to incorporate spatial information. 
The formulation of sIBP is also related to the stick-breaking representation of the Dirichlet process, and there are several works on introducing latent distributions for the transformed proportions \citep[e.g.][]{rodriguez2010latent,ren2011logistic,grazian2024spatio}.
However, the posterior distribution of the sIBP model is considerably different from those under the Dirichlet process model, so the existing sampling techniques for the Dirichlet process cannot be directly incorporated into 
our sIBP model. 

This paper is organized as follows. 
Section~\ref{sec:sIBP} introduces the proposed sIBP and discusses its theoretical properties.
In Section~\ref{sec:GS}, an efficient posterior computation algorithm for general latent variable models with sIBP is discussed. 
The numerical performance is demonstrated using simulation experiments in Section~\ref{sec:sim} and applications to two types of datasets in Section~\ref{sec:app}.
Technical proofs are provided in the appendix. 

%----------------------------------------------------------%
%        spatially dependent Indian Buffet Process         %
%----------------------------------------------------------%
\section{Spatially dependent Indian Buffet Process}
\label{sec:sIBP}

\subsection{Indian buffet process and stick-breaking representations}
Let $Z$ be a binary feature matrix, with $n$ rows for subjects and an unbounded number of columns for features, where each element $z_{ik}\in\{0,1\}$ indicates whether the $i$'th subject possess the $k$'th latent feature (i.e., $z_{ik}=1$) or not (Figure~\ref{figure:ibp}(b)).
In the Indian buffet metaphor, the rows of $Z$ represent customers, and the columns represent dishes, with $Z$ being constructed through a sequential sampling process. 
The first customer enters the restaurant and selects ${\rm Po}(\alpha)$ number of dishes,
where ${\rm Po}(\alpha)$ is a Poisson distribution with parameter $\alpha$.
Subsequently, $i$'th customer ($i>1$) samples each dish that has been previously taken with a probability of $m_k/i$, where $m_k$ denotes the number of customers who have chosen dish $k$ so far. 
Additionally, the $i$'th customer selects new dishes based on ${\rm Po}(\alpha/i)$. The selection of dishes by each customer can be represented in a binary feature allocation matrix $Z$.
Such process is called an {\it Indian Buffet Process} (IBP) (\cite{griffiths05ibp,griffiths2011indian}) 
from a metaphor of Indian restaurants prevalent in London,
and is denoted by ${\rm IBP}(\alpha)$.

As shown in \cite{griffiths2011indian}, the IBP can also be defined as the limit of a Beta--Bernoulli model.
Here, we consider the following latent variable model with finite $K$ features: 
\begin{equation}\label{eq:IBP}
z_{ik}\sim {\rm Bernoulli}\,(b_{k}), \quad b_k\sim {\rm Beta}\left(\frac{\alpha}{K}, 1\right),
\end{equation}
with a parameter $\alpha$, where the $b_k$ are generated independently and each $z_{ik}$ is independent of all other assignments under given $b_k$.
\cite{griffiths2011indian} showed that the distribution of $Z$ under $K\to\infty$ is equivalent to the one obtained by IBP. 
The IBP has several desirable properties:
in particular, the expectation of the number of features possessed by each subject under $K\to\infty$ is $\alpha$, which indicates the sparsity of feature selection. 
As an alternative representation of the Beta--Bernoulli model of (\ref{eq:IBP}), \cite{teh2007stick} introduced the stick-breaking representation of (\ref{eq:IBP}) under finite $K$, given by 
\begin{equation*}
z_{ik}\sim {\rm Bernoulli}\,(b_{k}), \quad b_{k}=\prod_{j=1}^k\, v_j, \ \ \ k=1,\ldots, K,
\end{equation*}
where $v_j\sim {\rm Beta}(\alpha, 1)$.
The above representation guarantees that the sequence $b_1,\ldots,b_K$ is strictly decreasing, and has the same distribution as IBP under $K\to\infty$.
We leverage this representation to introduce spatial correlation in the subsequent section. 
In what follows, we define the proposed process on an infinite sequence of latent features.
For posterior computation, we use a finite truncation level $K$, as is commonly done for
stick-breaking representations.

\subsection{Spatially dependent IBP}

\begin{figure}[t]
 \centering
 \includegraphics[width=0.47\textwidth]{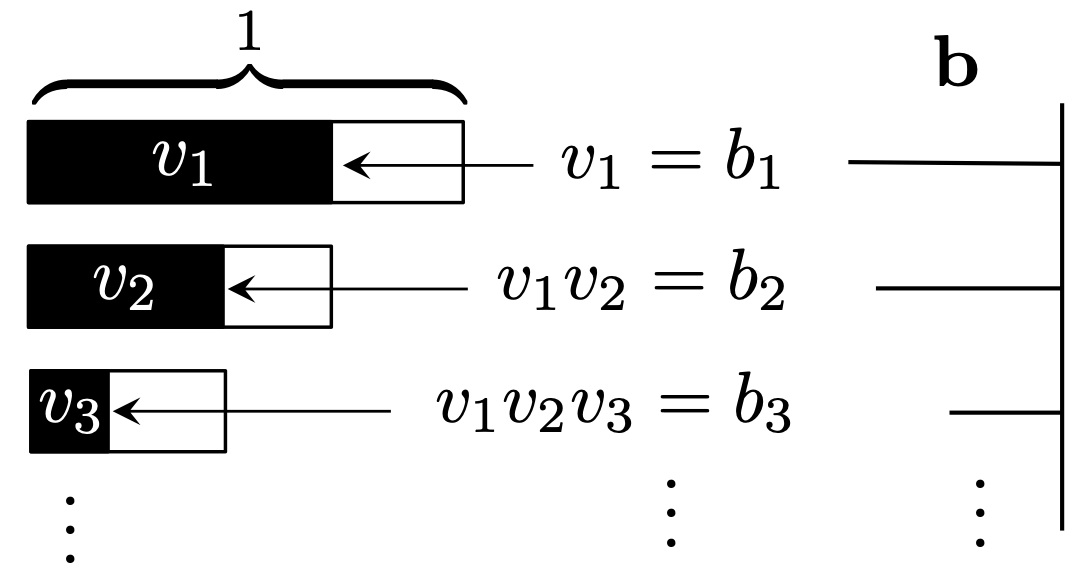}
 \caption{Stick-breaking representation of original IBP. Here, each breaking proportion $v_j$ is drawn from 
 Beta distribution $\mathrm{Beta}(\alpha,1)$.}
 \label{figure:ibp-sbp}
\end{figure}

To take account of spatial information, let $z_k(s)$ be a binary random variable at 
a geographic location $s$ and consider the following model: 
\begin{equation}\label{Sp-IBP-process}
z_k(s)\sim {\rm Bernoulli}\,(b_k(s)), \quad b_k(s)=\prod_{j=1}^k\, \sigma(u_j(s)), \ \ \ k=1,\ldots, K, 
\end{equation}
where $\sigma(x)=1/(1\!+\!e^{-x})$ is a logistic function and $u_j(s)$ follows a Gaussian process, independently for $j=1,\ldots,K$. 
We call {\it Spatially dependent Indian buffet process} (sIBP) of the process (\ref{Sp-IBP-process}). 
The process (\ref{Sp-IBP-process}) can be regarded as a spatial extension of the logistic stick-breaking process \citep{ren2011logistic}.
For $n$ locations denoted by $s_1,\ldots,s_n$, the joint model for $z_{ik}\equiv z_k(s_i) \ (i=1,\ldots,n)$ is described as 
\begin{equation}\label{Sp-IBP}
z_{ik}\sim {\rm Bernoulli}\,(b_{ik}), \quad b_{ik}=\prod_{j=1}^k\, \sigma(u_{ij}), \ \ \ k=1,\ldots, K.
\end{equation}
Here $u_{ik}$ is a continuous latent variable that follows a Gaussian process, with joint distribution
\begin{equation}\label{eq:GP}
u_k\equiv (u_{1k},\ldots,u_{nk})\sim {\cal N}(\mu 1_n, \tau^{-1}Q(\psi))
\end{equation}
independently for $k=1,\ldots,K$, where $\mu$ is a location parameter, $1_n$ is an $n$-dimensional vector of $1$, $\tau$ is a precision parameter and $Q(\psi)$ is a correlation matrix parametrized by $\psi$.
One example of $Q(\psi)$ is that its $(i, j)$-element is $\rho(d_{ij}; \psi)$, where $d_{ij}$ denotes some distance between $i$'th and $j$'th locations.
Standard choices are the Mat\'ern function, $\rho(d; \psi)=\{2^{\kappa-1}\Gamma(\kappa)\}^{-1}(d/\phi)^\kappa K_\kappa(d/\phi)$ with $\psi=(\phi, \kappa)$ and the exponential function, $\rho(d;\psi)=\exp(-d/\phi)$, where $\phi>0$ is a spatial range parameter and $\kappa>0$ is a smoothness parameter. 
Here $K_{\kappa}(\cdot)$ denotes the modified Bessel function of the third kind of order $\kappa$.
We call {\it Spatially dependent Indian buffet process} (sIBP) of the model (\ref{Sp-IBP}) with spatially correlated latent variables $\{u_{ik}\}$.
It should be noted that when the spatial range parameter is very large, the covariance matrix $Q(\psi)$ is degenerate and it holds that $u_{1k},\ldots,u_{nk}$ are identical with probability one and follow ${\cal N}(\mu, \tau^{-1})$. 
This corresponds to the stick-breaking representation of IBP with independent distributional assumptions for $v_k$. 

Although the construction in (\ref{Sp-IBP-process}) is introduced with a
finite truncation level $K$, it naturally extends to the infinite-feature
case by letting $K\to\infty$. 
By Kolmogorov's extension theorem, this
infinite-feature construction defines a valid spatial stochastic process.
This is summarized in the following proposition; the proof is provided in
the Supplementary Material.

\begin{prp}
The infinite-feature version of (\ref{Sp-IBP-process}), obtained by
letting $K\to\infty$, defines a well-defined spatial stochastic process
$\{Z(s):s\in\mathcal{S}\}$, where
$Z(s)=(z_1(s),z_2(s),\ldots)$.
\end{prp}

\subsection{Properties of sIBP}

We present some theoretical properties of sIBP.
First, we prove that the expected number of non-zero entries is finite in the limiting case. 

\begin{prp}\label{prp1}
The binary matrix $Z$ generated by sIBP is sparse, that is, $\mathbb{P}(z_{ik}=1)\to 0$ as $k\to\infty$. 
Further, let $c_i=\sum_{j=1}^k z_{ij}$ be the number of features that the $i$th subject possess. 
Then, $\lim_{k\to\infty} E[c_i]=\delta_1/(1-\delta_1)$ and
$$
\lim_{k\to\infty} {\rm Var}(c_i)=
\frac{\delta_1}{1-\delta_1} \left(1+\frac{2\delta_2}{1-\delta_2}-\frac{\delta_1}{1-\delta_1}\right)
$$
for each $i=1,\ldots,n$, where 
$$
\delta_p=\int_{-\infty}^{\infty} \sigma(x)^p\phi(x; \mu, \tau^{-1})dx, \ \ \ \ p=1,2
$$
are constants dependent on $\mu$ and $\tau$. 
\end{prp}

The Proposition~\ref{prp1} ensures that the number of sampled features is finite, namely, $\mathbb{P}(c_i<\infty)=1$.
Furthermore, it can be seen that the expectation of the total number of non-zero elements is $n\delta_1/(1-\delta_1)$.
Further, when $\mu$ and $\tau$ are chosen such that $2\delta_2/(1-\delta_2)=\delta_1/(1-\delta_1)$, the limit of $\mathbb{E}[c_i]$ and ${\rm Var}(c_i)$ are the same as the standard IBP. 
It should also be noted that the spatial range parameter $\psi$ (controlling the correlation given a fixed distance) of the associated Gaussian process does not appear in the formula.
This indicates that the basic performance of sIBP depends only on the parameters, $\mu$ and $\tau$, of the marginal distributions.

We next investigate how the spatial correlation is introduced in the generated binary matrices through the latent Gaussian process.
To this end, we evaluate the joint probability that two binary variables are both 1, as given in the following proposition:

%   Proposition
\begin{prp}\label{prp3}
Under the sIBP model, $\mathbb{P}(z_{ik}=1, z_{jk}=1)=D_{\mu,\tau}(\rho_{ij})^k$ for each $i,j=1,\ldots,n$, where $D_{\mu,\tau}(\rho_{ij})=\mathbb{E}[\sigma(u_{ik})\sigma(u_{jk})]$ and the expectation is taken with respect to 
$(u_{ik},u_{jk})\sim N_2((\mu,\mu), \Sigma)$ with $\Sigma_{11}=\Sigma_{22}=1/\tau$, 
$\Sigma_{12}=\Sigma_{21}=\rho_{ij}/\tau$ and $\rho_{ij}=\rho(\|s_i-s_{j}\|; \psi)$.
Furthermore, $D_{\mu,\tau}(\rho)$ is an increasing function of $\rho\in (0,1)$. 
\end{prp}

Proposition~\ref{prp3} indicates that the probability that the two subjects simultaneously possess the $j$'th
feature is dictated by spatial correlation $\rho_{ij}$ and spatial range parameter $\psi$.

\begin{figure}[htbp!]
\centering
\includegraphics[width=\linewidth]{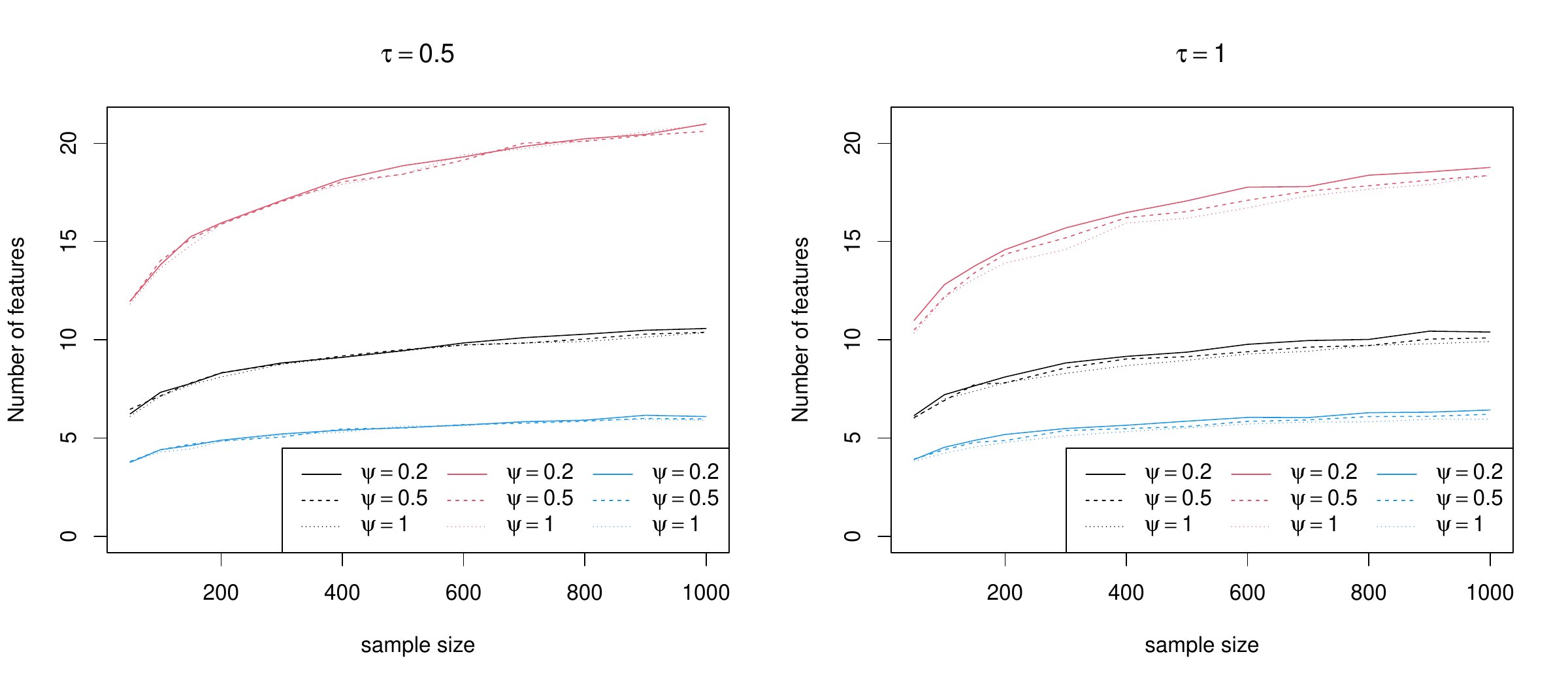}
\caption{Number of common factors under sIBP with $\mu=-1$ (blue), $\mu=0$ (black), and $\mu=1$ (red), as a function of sample size. 
}
\label{fig:feature}
\end{figure}

We next consider the number of (non-null) common features, defined as $K^{\ast}=\sum_{k=1}^K I(c_k>0)$, where $c_k=\sum_{i=1}^n z_{ik}$.
Then, it holds that 
\begin{align*}
\mathbb{E}[K^{\ast}]=\sum_{k=1}^K \mathbb{P}(c_k>0)
&=\sum_{k=1}^K \bigg\{1 - \mathbb{P}(z_{1k}=0, \ldots, z_{nk}=0)\bigg\} \\
&=\sum_{k=1}^K \bigg\{1 - \mathbb{E}\bigg[\prod_{i=1}^n \Big(1-\prod_{j=1}^k \sigma(u_{ij})\Big)\bigg]\bigg\},
\end{align*}
where the expectation is taken with respect to the joint distribution of $u_k\sim {\cal N}(\mu 1_n, \tau^{-1}Q(\psi))$.
Although the above expectation cannot be expressed in a closed form, we simulated the expectations by Monte Carlo integration under various choices of $(\mu, \tau, \psi)$.
Specifically, we consider the exponential covariance function, $Q(\psi)_{ij}=\exp(-\|s_i-s_j\|/\psi)$, and generated the location information from the uniform distribution on $[0,1]\times [0,1]$.
Regarding the parameter values, we consider all the combinations of $\mu\in \{-1, 0, 1\}$, $\tau\in \{0.5, 1\}$ and $\psi\in \{0.2, 0.5, 1\}$.
In Figure~\ref{fig:feature}, we show $\mathbb{E}[K^{\ast}]$ as a function of the sample size. 
The results indicate that the parameter $\mu$ significantly controls the number of features, that is, larger value of $\mu$ leads to larger number of features in the prior distribution. 
On the other hand, $\tau$ and $\psi$ have little effect on the number of features.

% Figure: number of features 

%----------------------------------------------------------%
%                  Implementation                          %
%----------------------------------------------------------%
\section{Implementation of sIBP}

\subsection{Latent variable models with sIBP}

Let $x_1,\ldots,x_n$ be an observed data, and $s_i \ (i=1,\ldots,n)$ be a location information (e.g., longitude and latitude) associated with $x_i$.  
We consider the following latent variable model: 
\begin{equation}\label{Model}
\begin{split}
&x_i\sim f\left(x_i\mid\{z_{ik}\}_{k=1}^K;\{\theta_k\}_{k=1}^K\right), \quad
z_{ik}\sim {\rm Bernoulli}\Big(\prod_{j=1}^k \sigma(u_{ij})\Big), \ \ \ 
i=1,\ldots,n,\\
& (u_{1k},\ldots,u_{nk})\sim N(\mu 1_n, \tau^{-1}Q(\psi)), \ \ \ \ k=1,\ldots,K,
\end{split}
\end{equation}
where $f(x_i\mid\{z_{ik}\}_{k=1}^K;\{\theta_k\}_{k=1}^K)$ is a density (or probability) function of $x_i$.
Note that $\theta_k$ is related to the distribution of $x_i$, and $(u_{1k},\ldots,u_{nk})$ are independent for $k=1,\ldots,K$. 
The unknown parameters in the latent variable model (\ref{Model}) are $\theta_k$, $\tau$, and $\psi$, for which we assign prior distributions for Bayesian inference. 
In particular, we employ a conditionally conjugate prior, $\tau\sim {\rm Ga}(a_{\tau}, b_{\tau})$, $\mu\sim {\rm N}(m_{\mu}, S_{\mu})$ and gamma priors for each element of $\psi$. 
Regarding the prior for $\theta_k$, we use a class of repulsive prior \citep[e.g.][]{petralia2012repulsive,xie2020bayesian} instead of assuming independent priors for $\theta_1,\ldots,\theta_K$ to prevent redundant expression of the latent factors. 
Let $\pi(\theta_k)$ be a marginal prior of $\theta_k$.
Then, the repulsive prior for $\theta_1,\ldots,\theta_K$ is given by 
$$
\pi(\theta_1,\ldots,\theta_K)=\frac{1}{Z_K}\left\{\prod_{k=1}^K \pi(\theta_k)\right\} \min_{1\leq k<k'\leq K} g(\|\theta_k-\theta_{k'}\|),
$$
where $Z_K$ is a normalizing constant and $g: \mathcal{R}_{+}\to [0,1]$ is a strictly increasing function with $g(0)=0$. 
The repulsive prior eliminates the prior mass from the region where $\theta_k$ and $\theta_{k'}$ are close.
For the specific choice of $g(\cdot)$, we use $g(x)=I(x>\delta)(1-\delta/x)$ with a tuning constant $\delta>0$.
Since the results would not be sensitive to the choice of $\delta$ as long as $\delta$ is small, we set $\delta=10^{-3}$ throughout the numerical studies in this paper. 

\subsection{Posterior computation via Gibbs sampler}\label{sec:GS}

For a general latent variable model with sIBP (\ref{Model}), we here provide a Gibbs sampler to simulate the posterior distributions by using a novel combination of binomial expansion and the P\'olya-gamma data augmentation \citep{polson2013bayesian} for logistic regression.
Define $Z$, $U$ and $\Theta$ be sets of $z_{ik}$, $u_{ik}$ and $\theta_k$, respectively.
Under the prior for these parameters, the joint posterior distribution $\pi(Z, U, \Theta, \tau, \psi|D)$ of the latent variables ($Z$ and $U$) and unknown parameters ($\Theta, \tau$ and $\psi$) given a set of observed data $D=\{x_1,\ldots,x_n\}$ can be expressed as 
\begin{align*}
\pi(Z, U, \Theta, \tau, \psi | D) \propto&
\prod_{i=1}^n
f(x_i| \{z_{ik}\}_{k=1}^K; \{\theta_k\}_{k=1}^K)
\prod_{k=1}^K
\left\{
\prod_{j=1}^k \sigma(u_{ij})
\right\}^{z_{ik}}
\left\{
1-\prod_{j=1}^k \sigma(u_{ij})
\right\}^{1-z_{ik}} \\
&\times p(\tau)p(\psi)\pi(\theta_1,\ldots,\theta_K)
\prod_{k=1}^K
\phi(u_k; \mu 1_n, \tau^{-1}Q(\psi)).
\end{align*}
We use Gibbs sampling to generate random samples from the above posterior distribution. 
The details are described as follows:

\paragraph{Update of $u_k$:}
For each $i$ and $k$, the full conditional distribution of $u_k$ is proportional to 
\begin{align*}
&\phi(u_k; \mu 1_n, \tau^{-1}Q(\psi))\prod_{i=1}^n \prod_{j=k}^K\sigma(u_{ik})^{z_{ij}}\bigg\{1-\sigma(u_{ik})\prod_{h=1, h\neq k}^{j}\sigma(u_{ih})\bigg\}^{1-z_{ij}}\\
=&\ \phi(u_k; \mu 1_n, \tau^{-1}Q(\psi))\prod_{i=1}^n \prod_{j=k}^K \frac{(e^{u_{ik}})^{z_{ij}}(1+C_{ikj}e^{u_{ik}})^{1-z_{ij}}}{1+e^{u_{ik}}},
\end{align*}
where $C_{ikj}=1-\prod_{h=1,h\neq k}^j \sigma(u_{ih})\in (0,1)$.  For the term 
$(1+C_{ikj}e^{u_{ik}})^{1-z_{ij}}$, we use the binomial expansion $(1+C_{ikj}e^{u_{ik}})^{1-z_{ij}}=\sum_{s_{ikj}=0}^{1-z_{ij}}
\left(C_{ikj}e^{u_{ik}}\right)^{s_{ikj}}$.
Here $s_{ikj}\in\{0,1\}$ if $z_{ij}=0$, while $s_{ikj}=0$ if $z_{ij}=1$.
Then, the likelihood term can be augmented as
\begin{align*}
\prod_{j=k}^K\frac{(e^{u_{ik}})^{z_{ij}}(e^{u_{ik}})^{(1-z_{ij})s_{ikj}}}{1+e^{u_{ik}}}=\frac{(e^{u_{ik}})^{\sum_{j=k}^K\{z_{ij}+(1-z_{ij})s_{ikj}\}}}{(1+e^{u_{ik}})^{K-k+1}}.
\end{align*}
Note that the full conditional probability being $s_{ikj}=1$ for $z_{ij}=0$ is $C_{ikj}e^{u_{ik}}/(C_{ikj}e^{u_{ik}}+e^{u_{ik}z_{ij}})$, whereas $s_{ikj}=0$ with probability one when $z_{ij}=1$.
Then, applying the P\'olya-gamma data augmentation \citep{polson2013bayesian}, the above function can be expressed as 
$$
\frac{ (e^{u_{ik}})^{\sum_{j=k}^K \{z_{ij}+(1-z_{ij})s_{ikj}\}} }{ (1+e^{u_{ik}})^{K-k+1} }
\propto \exp(\kappa_{ik}u_{ik})\int_0^{\infty} \exp\left(-\frac12\omega_{ik}u_{ik}^2\right)f_{\rm PG}(\omega_{ik}|K-k+1, 0)d\omega_{ik},
$$
where $f_{\rm PG}(\cdot|b, c)$ is the density function of P\'olya-gamma distribution
${\rm PG}(b,c)$ and 
$$
\kappa_{ik}=\sum_{j=k}^K \{z_{ij}+(1-z_{ij})s_{ikj}\}-\frac12 (K-k+1).
$$
Note that the full conditional distribution of $\omega_{ik}$ is ${\rm PG}(K-k+1,u_{ik})$.
Using this augmentation, the full conditional distribution of $u_k$ becomes
${\cal N}(A_k^{-1}B_k, A_k^{-1})$, where 
$$
A_k={\rm diag}(\omega_{1k},\ldots,\omega_{nk})+\tau Q(\psi)^{-1}, \ \ \ B_k=(\kappa_{1k},\ldots,\kappa_{nk})^\top + \tau Q(\psi)^{-1} 1_n \mu \,.
$$

\paragraph{Update of $z_{i1}, \ldots, z_{iK}$:} For each $i$, define $b_{ik}=\prod_{j=1}^k \sigma(u_{ij})$.
Then the full conditional distribution of $(z_{i1},\ldots,z_{iK})$ is proportional to
$$
f\left(x_i \mid \{z_{ik}\}_{k=1}^K; \{\theta_k\}_{k=1}^K\right)
\prod_{k=1}^K
b_{ik}^{z_{ik}}(1-b_{ik})^{1-z_{ik}}.
$$
When $K$ is not large, we can compute the above unnormalized probability on
$(z_{i1},\ldots,z_{iK})\in\{0,1\}^K$, and update the binary vector
$(z_{i1},\ldots,z_{iK})$ simultaneously. When $K$ is large, one can update
$z_{ik}$ separately by computing the full conditional probability of $z_{ik}=1$.

\medskip

\paragraph{Update of $\theta_{k}$:}
For each $k$, the full conditional of $\theta_k$ is proportional to 
\begin{align*}
\pi(\theta_k) \prod_{i=1}^n f\big(x_i\mid\{z_{ij}\}_{j=1}^K;\{\theta_j\}_{j=1}^K\big)\min_{1\leq k'\leq K, k'\neq k} g(\|\theta_k-\theta_{k'}\|),
\end{align*}
and the specific sampling algorithms depend on the distributional assumption for $x_i$. 
A typical strategy is to generate a proposal $\theta_k^{\ast}$ from a distribution proportional to $\pi(\theta_k) \prod_{i=1}^n f\big(x_i\mid\{z_{ij}\}_{j=1}^K;\{\theta_j\}_{j=1}^K\big)$ and accept it with probability 
$$
\min_{1\leq k'\leq K, k'\neq k} g(\|\theta_k^{\ast}-\theta_{k'}\|)/\min_{1\leq k'\leq K, k'\neq k} g(\|\theta_k^{\dagger}-\theta_{k'}\|),
$$
for the current value $\theta_k^{\dagger}$. 

\paragraph{Update of $\tau$:}
The full conditional of $\tau$ is ${\rm Ga}(\tilde{a}_{\tau}, \tilde{b}_\tau)$, where $\tilde{a}_{\tau}=a_{\tau}+\frac12nK$ and $\tilde{b}_\tau=b_{\tau}+\frac12\sum_{k=1}^K(u_k-\mu 1_n)^{\top}Q(\psi)^{-1}(u_k-\mu 1_n)$.

\paragraph{Update of $\mu$:}
The full conditional of $\mu$ is ${\cal N}(A_{\mu}^{-1}B_{\mu}, A_{\mu}^{-1})$, where $A_{\mu}=K\tau 1_n^{\top} Q(\psi)^{-1} 1_n + S_\mu^{-1}$ and $B_{\mu}=\tau  1_n^{\top} Q(\psi)^{-1}\sum_{k=1}^Ku_k + S_\mu^{-1}m_{\mu}$.

\paragraph{Update of $\psi$:}
The full conditional of $\psi$ is proportional to 
$$
\pi(\psi)|Q(\psi)|^{-K/2}\exp\left\{-\frac{\tau}{2}\sum_{k=1}^K (u_k-\mu 1_n)^{\top}Q(\psi)^{-1}(u_k-\mu 1_n)\right\}, 
$$
for which we use a random-walk Metropolis-Hastings algorithm.

\medskip
The above Gibbs sampling algorithm can be easily implemented since most of the full conditional distributions are familiar due to the binomial expansion and P\'olya-gamma data augmentation.

Owing to the Gaussian process assumption, we can predict latent binary vectors in (arbitrary) unobserved locations. 
Let $s$ be an unobserved location of interest. 
Then, the conditional distribution of $u_k(s)$ given $u_k=(u_{1k},\ldots,u_{nk})$ is ${\cal N}(\mu+C(s)Q(\psi)^{-1}(u_k-\mu 1_n), \tau^{-1}-\tau^{-1}C(s)Q(\psi)^{-1}C(s))$, where $C(s)=(\rho(d(s, s_1); \psi),\ldots,\rho(d(s, s_n); \psi))$ and $d(s, s_1)$ is a distance between unobserved location $s$ and observed location $s_i$. 
Hence, one can easily simulate the binary indicator $z_k(s)$ at site $s$ based on the posterior predictive distribution of $b_k(s)\equiv \prod_{j=1}^k \sigma(u_j(s))$. 

\subsection{Scalable implementation of sIBP}\label{sec:scalable}

When the number of locations, $n$, is large, the posterior computational cost may be intensive due to the Gaussian process assumption for the latent variable $u_{ik}$.
To make the proposed method scalable under a large number of locations, we adopt the nearest-neighbor Gaussian process \citep{datta2016hierarchical}.
Instead of the Gaussian process model in (\ref{eq:GP}), 
we employ the following nearest-neighbor Gaussian process approximation:
\begin{align*}
\pi(u_{1k},\ldots,u_{nk}) = \prod_{i=1}^n \phi(u_{ik}; \mu+B_i\{u_k(N(s_i))-\mu 1_{m_i}\}, \tau^{-1}F_i), \quad k=1,\ldots,K,
\end{align*}
where $N(s_i)$ is the set of $m_i$ nearest neighbors of the location $s_i$, 
$u_k(N(s_i))$ is a sub-vector of $u_k$ collecting random variables on $N(s_i)$, and
$$
B_i = Q_{s_i,N(s_i)}(\psi)Q_{N(s_i)}(\psi)^{-1}, 
\quad
F_i = 1 - Q_{s_i,N(s_i)}(\psi)Q_{N(s_i)}(\psi)^{-1}Q_{N(s_i),s_i}(\psi).
$$
Here $Q_{s_i,N(s_i)}(\psi)\equiv {\rm Cor}\{u_{ik},u_k(N(s_i))\}$ and 
$Q_{N(s_i)}(\psi)\equiv {\rm Cor}\{u_k(N(s_i)),u_k(N(s_i))\}$ are respectively 
a sub-vector and sub-matrix of $Q(\psi)$. Under the model, the full conditional 
distribution of $u_{ik}$ is also normal as in the standard Gaussian process, so we can 
use a similar Gibbs sampling algorithm under the nearest-neighbor Gaussian process.

\subsection{Example 1: Spatial factor models for large-dimensional multinomial data}\label{sec:multinomial-model}

To reveal the differences in the geographical distribution of dialects for various meanings, as considered in Section~\ref{sec:dialect}, we need to deal with large-dimensional multinomial data.
Let $x_i=(x_{i1},\ldots,x_{iM})$ be an $M$-dimensional vector of multinomial observations, where $x_{im}$ takes one of $\{1,\ldots,c_m\}$ for $m=1,\ldots,M$ and $i=1,\ldots,n$.
Here $n$ is the number of areas and $M$ is the number of meanings, which could be large to achieve appropriate characterization of each dialect. 
Further, let $\pi_{iml} \ (l=1,\ldots,c_m)$ be a multinomial probability, so that $x_{im}$ is generated from the categorical distribution on $\{1,\ldots,c_m\}$ with probability $(\pi_{im1},\ldots,\pi_{imc_m})$.

In this setting, we are interested in low-dimensional representations of the difference of dialects. 
To this end, we model the probability as 
$$
\pi_{iml}=\frac{\exp\big(\eta_{ml}+\sum_{k=1}^K z_{ik}\theta_{kml}\big)}{\sum_{l'=1}^{c_m} \exp\big(\eta_{ml'}+\sum_{k=1}^K z_{ik}\theta_{kml'}\big)}, \ \ \ \ l=1,\ldots,c_m, 
$$
where $(\eta_{m1},\ldots,\eta_{mc_m})$ is a baseline common to all the areas and $\theta_{kml}$ is a continuous value of $k$th factor. 
We also set $\eta_{m1}=0$ and $\theta_{km1}=0$ for identifiability. 
We further assume that $\eta_{ml}\sim {\cal N}(0, \gamma_0^{-1})$ and $\theta_{kml}\sim {\cal N}(0, \gamma_k^{-1})$ independently for $m=1,\ldots,M$ and $l=2,\ldots,c_m$.
Note that the distribution of $x_i$ is 
$$
f\big(x_i\mid\{z_{ik}\}_{k=1}^K;\{\theta_k\}_{k=1}^K, \eta\big)=\prod_{m=1}^M\prod_{l=1}^{c_m}
\left[\frac{\exp\big(\eta_{ml}+\sum_{k=1}^K z_{ik}\theta_{kml}\big)}{\sum_{l'=1}^{c_m} \exp\big(\eta_{ml'}+\sum_{k=1}^K z_{ik}\theta_{kml'}\big)}\right]^{I(x_{im}=l)},
$$
so that we can generate $z_i$, $u_k$, $\tau$ and $\psi$ through their full conditional distributions as given in Section~\ref{sec:GS}.
On the other hand, the full conditional of $\Theta_{ml}=(\eta_{ml}, \theta_{1ml},\ldots,\theta_{Kml})$ is 
\begin{align*}
&\phi_{K+1}(\Theta_{ml}; 0, \Gamma^{-1})\prod_{i=1}^n
\frac{\exp\big(\eta_{ml}+\sum_{k=1}^K z_{ik}\theta_{kml}\big)^{I(x_{im}=l)}}{\sum_{l'=1}^{c_m} \exp\big(\eta_{ml'}+\sum_{k=1}^K z_{ik}\theta_{kml'}\big)}\\
\propto \ 
& 
\phi_{K+1}(\Theta_{ml}; 0, \Gamma^{-1})
\prod_{i=1}^n \frac{\exp\big(z_i^\top \Theta_{ml} -C_{iml}\big)^{I(x_{im}=l)}}{1+\exp\big(z_i^\top \Theta_{ml} - C_{iml}\big)},
\end{align*}
where $\Gamma={\rm diag}(\gamma_0,\gamma_1,\ldots,\gamma_K)$, $z_i=(1, z_{i1},\ldots,z_{iK})$ and $C_{iml}=\log\sum_{l'=1, l'\neq l}^{c_m} \exp\big(z_i^\top \Theta_{ml'})$.
Using the P\'olya-gamma data augmentation \citep{polson2013bayesian}, the above distribution can be augmented as 
$$
\phi_{K+1}(\Theta_{ml}; 0, \Gamma^{-1})
\prod_{i=1}^n
\exp\Big\{\kappa_{iml}(z_i^\top \Theta_{ml}-C_{iml})-\frac12\omega_{ml}(z_i^\top \Theta_{ml}-C_{iml})^2\Big\}f_{\rm PG}(\omega_{iml}; 1, 0),
$$
where $\kappa_{iml}=I(x_{im}=l)-1/2$.
The full conditional of $\omega_{iml}$ is ${\rm PG}(1, z_i^\top \Theta_{ml}-C_{iml})$, and that of $\Theta_{ml}$ is ${\cal N}(A_{ml}^{-1}B_{ml}, A_{ml}^{-1})$, where 
$$
A_{ml}=\sum_{i=1}^n\omega_{iml}z_iz_i^\top  + \Gamma, \ \ \ \ B_{ml}=\sum_{i=1}^n z_i(\kappa_{iml}+\omega_{iml}C_{iml}).
$$

\subsection{Example 2: Spatial factor models for multivariate count data}\label{sec:count-model}
To investigate the vegetation of multiple species of trees in Japan, considered in Section~\ref{sec:tree}, we want to extract spatial factors regarding the geographical distributions of multiple species of trees. 
To this end, we consider the case with multivariate count responses. 
Let $x_i=(x_{i1},\ldots,x_{iM})$ be an $M$-dimensional vector of count observations. 
We assume that each element $x_{im}$ is drawn from a negative Binomial distribution:
$$
x_{im}\sim {\rm NB}(\lambda_{im}, \nu_m), \ \ \ \lambda_{im}=\exp\left(\eta_{m}+\sum_{k=1}^K z_{ik}\theta_{km}\right),
$$
where $\nu_m$ is a dispersion parameter, $\eta_m$ is a baseline term and $(\theta_{k1},\ldots,\theta_{kM})$ is a vector of mean parameters for the $k$th factor.
Note that the distribution of $x_i$ is 
$$
f(x_i\mid \{z_{ik}\}_{k=1}^K; \{\theta_k\}_{k=1}^K,\eta)=\prod_{m=1}^M \frac{\Gamma(\nu_m+x_{im})}{\Gamma(\nu_m)x_{im}!}\left(\frac{\nu_m}{\nu_m+\lambda_{im}}\right)^{\nu_m}\left(\frac{\lambda_{im}}{\nu_m+\lambda_{im}}\right)^{x_{im}}
$$
so that $\mathbb{E}[x_{im}]=\lambda_{im}$ and ${\rm Var}(x_{im})=\lambda_{im}+\lambda_{im}^2/\nu_m$.

We assign prior distributions, $\nu\sim {\rm Ga}(a_{\nu}, b_{\nu})$, $\eta_{m}\sim {\cal N}(0, \gamma_0^{-1})$ and $\theta_{km}\sim {\cal N}(0, \gamma_k^{-1})$ independently for $m=1,\ldots,M$.
The sampling algorithm for parameters other than $\nu_m, \eta_m$ and $\theta_{km}$ is the same as in Section~\ref{sec:GS}.
The detailed sampling algorithm for these parameters is given as follows: 

\paragraph{Sampling from $\Theta_m=(\eta_m, \theta_{1m},\ldots,\theta_{Km})$:}
Using the P\'olya-gamma data augmentation, we first generate latent variable $\omega_{im}$ from ${\rm PG}(x_{im}+\nu_m, \eta_{m}+\sum_{k=1}^K z_{ik}\theta_{km}-\log \nu_m)$.
Then one can generate $\Theta_m$ from ${\cal N}(A_{m}^{-1}B_{m}, A_{m}^{-1})$, where 
$$
A_{m}=\sum_{i=1}^n\omega_{im}z_iz_i^\top  + \Gamma, \ \ \ \ B_{m}=\sum_{i=1}^n z_i(\kappa_{im}+\omega_{im}\log\nu_m).
$$
and $\kappa_{im}=(x_{im}-\nu_m)/2$.
Here $\Gamma={\rm diag}(\gamma_0,\gamma_1,\ldots,\gamma_K)$, $z_i=(1, z_{i1},\ldots,z_{iK})$.

\paragraph{Sampling from $\nu_m$:}
For $m=1\ldots,M$, we employ a random-walk Metropolis-Hastings algorithm.

%----------------------------------------------------------%
%                    Simulation                            %
%----------------------------------------------------------%
\section{Simulation Study}\label{sec:sim}

\subsection{Recovery of latent factors }

\begin{figure}[htbp!]
\centering
\includegraphics[width=\linewidth]{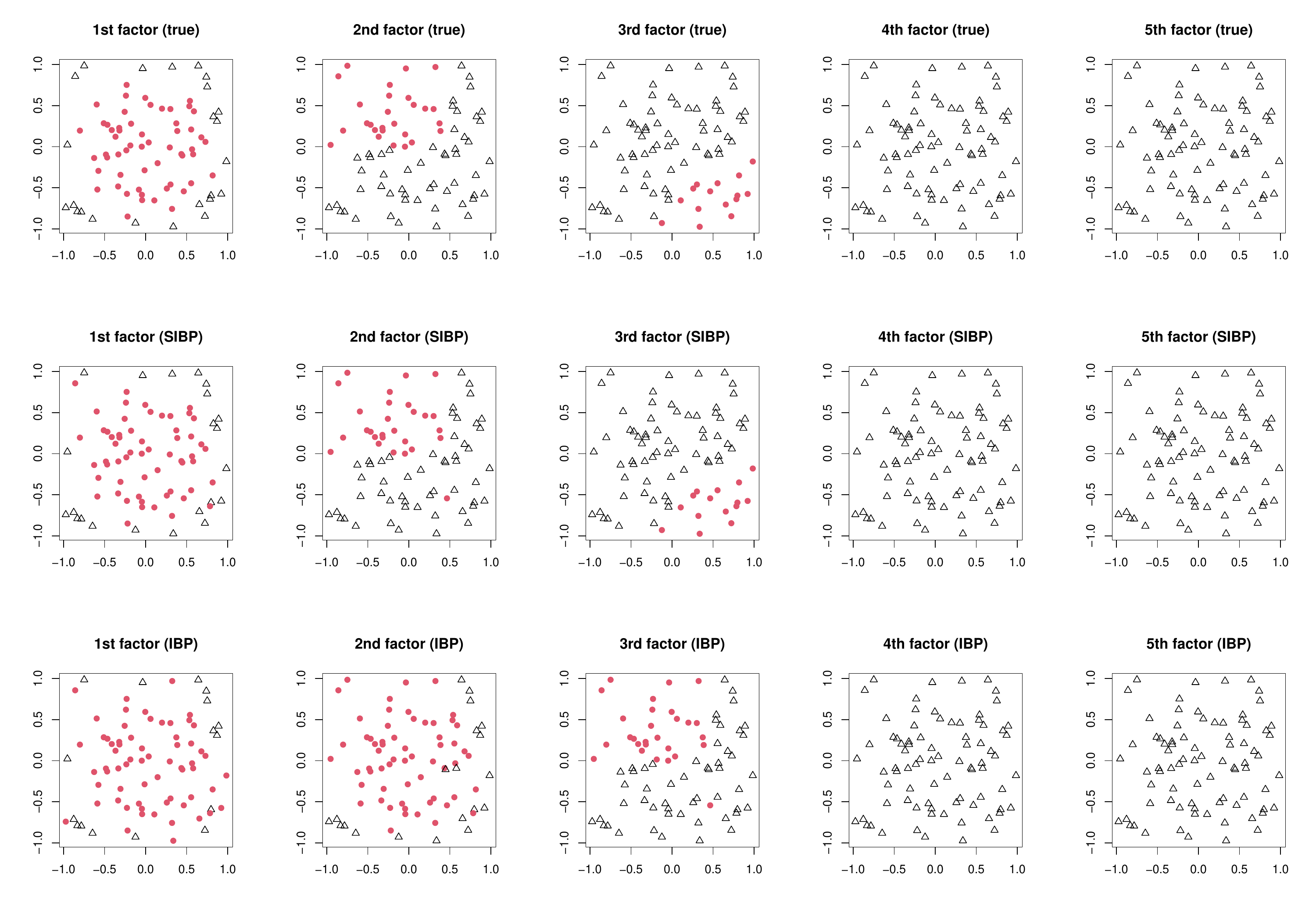}
\caption{The true latent factors (top) and estimated factors by the proposed sIBP (middle) and the standard IBP (bottom). 
}
\label{fig:sim-cluster1}
\end{figure}

We first demonstrate the numerical performance of the proposed sIBP through the multinomial models introduced in Section~\ref{sec:multinomial-model}.
For $i=1,\ldots,n$, we let $x_i=(x_{i1},\ldots,x_{iM})$ with be a observed vector, where $x_{im} \ (m=1,\ldots,M)$ independently follows a multinomial distribution on $\{1,\ldots,L\}$ with probability $\pi_{iml} \ (l=1,\ldots,L)$.
Further, let $s_i=(s_{i1}, s_{i2})$ be the two-dimensional vector of location information generated from the uniform distribution on $[-1, 1]\times [-1, 1]$.
We set $n=80$, $M=50$ and $L=5$ in this study.
We consider the following true structure of the multinomial probability $\pi_{iml}$:  
$$
\pi_{iml}=
 \frac{\exp \left(\sum_{k=1}^{K^{\ast}} z_{ik}\theta_{kml} \right)}
      {\sum_{l'=1}^L\exp \left(\sum_{k=1}^{K^{\ast}} z_{ik}\theta_{kml'} \right)},  
$$ 
where $z_{ik}$ is a binary variable and $\theta_{kml}$ determines the effects of the $k$'th latent factor. 
We adopt the latent factor defined as 
\begin{align*}
{\rm (I)}  \ \ \ \ &z_{i1}=I(s_{i1}^2+s_{i2}^2<(0.9)^2), \ \ \ \ z_{i2}=I(s_{i1}<0.5, s_{i2}>0), \ \ \ \ 
z_{i3}=I(s_{i1}<s_{i2}-0.7), \\
&\theta_{1ml}\sim {\cal N}(1,1), \ \ \ \ 
\theta_{2ml}\sim {\cal N}(0.5, (1.5)^2)), \ \ \ \ 
\theta_{3ml}\sim {\cal N}(0, 4),
\end{align*}
where $K^{\ast}=3$ as the true number of latent factors. 
The realized location information and true clustering structures are shown in the top row of Figure~\ref{fig:sim-cluster1}.
Note that $\sum_{i=1}^nz_{i1}=56$, $\sum_{i=1}^nz_{i2}=27$ and $\sum_{i=1}^nz_{i3}=15$.

Based on the generated data $(x_i, s_i)$, we fitted the multinomial factor model described in Section~\ref{sec:multinomial-model} with the proposed sIBP and the standard IBP.
We set $K=10$, the maximum number of factors in both sIBP and IBP models.
We generated 2,000 posterior samples after discarding the first 2,000 samples.
We then computed the posterior probability being $z_{ik}=1$, and we then detected areas having the posterior probability greater than $0.5$ for each $k=1,\ldots, K$.  
We found that only the first three factors are non-null, and the other four clusters are empty (i.e., $z_{ik}=0$ in all the areas) under both sIBP and IBP models, showing their adaptive property in selecting the number of latent factors. 
In Figure~\ref{fig:sim-cluster1}, we show the spatial distribution of detected factors. 
The results show that sIBP can recover the true latent structures almost perfectly, where the rand indices are 0.95 (1st factor), 0.95 (2nd factor), and 1 (3rd factor).
On the other hand, IBP fails to recover the latent factors, and particularly, the detected factors in the first and second factors are almost identical and not interpretable, possibly due to the less flexibility of IBP in handling spatial correlation. 

\subsection{Comparison of prediction performance}
We next compare the performance of sIBP with other existing statistical or machine learning methods in estimating the underlying true multinomial probabilities, under situations where the exact spatial factors do not necessarily exist. 
We consider $M=50$ dimensional multinomial observations (five items for each dimension) for $i=1,\ldots,n$, where each observation is equipped with location information uniformly distributed on $[-1,1]\times [-1,1]$. 
For the true multinomial probabilities, we consider the following two scenarios in addition to Scenario (I) given in the previous section: 
\begin{align*}
{\rm (II)}  \ \ \ \ &z_{i1}=I(s_{i1}<0), \ \ \ \ z_{i2}=I(|s_{i2}|>s_{i1}), \ \ \ \ 
z_{i3}=I(s_{i1}>0.5, s_{i2}<-0.5), \\
&\theta_{1ml}\sim {\cal N}(1,3), \ \ \ \ 
\theta_{2ml}\sim {\cal N}(0.5, (1.5)^2)), \ \ \ \ 
\theta_{3ml}\sim t_3(2, 1),\\
{\rm (III)}  \ \ \ \ & z_{ilm}=I(s_{i2}>c_{lm}s_{i1}), \ \ \ c_{lm}\sim U(-2,2), \ \ \ \ \pi_{iml}\propto z_{ilm}s_{i1}+(1-z_{ilm})(s_{i1}^2+s_{i2}^2),
\end{align*}
where $t_m(a,b)$ denotes the $t$-distribution with $m$-degrees of freedom having location and scale parameters, $a$ and $b$, respectively. 
Note that the multinomial probabilities in Scenario~(III) are expressed as a function of the two-dimensional location information. 
The whole region is divided into two sub-regions with different underlying structures of multinomial probabilities.

Based on the observations on $n$ locations, we predict the multinomial probabilities in $n_{\rm test}$ locations.
Let $\pi_{n+i,ml}$ be the true multinomial probability at the $i$th test location and $l$th category in the $m$th feature, and $\hat{\pi}_{iml}$ be the estimated one. 
The prediction performance is evaluated by the total mean squared errors (MSE), $(n_{\rm test}M)^{-1}\sum_{i=n+1}^{n+n_{\rm test}} \sum_{m=1}^M \sum_{l=1}^{c_m}(\hat{\pi}_{iml}-\pi_{iml})^2$. 
In this simulation study, we considered $(n, n_{\rm test})=(50, 20)$.
In addition to sIBP and IBP, we consider the following methods as competitors: 

\begin{description}
\item[Extreme gradient boosting tree (XGBoost):]
Using location information $(s_{i1},s_{i2})$ as input and $x_{im}$ as output, we employ the extreme gradient boosting tree (XGBoost) using the R package ``xgboost", separately for $m=1,\ldots,M$. 
\item[Multinomial Gaussian process (MnGP):]
For each $m$, the multinomial probability $\pi_{iml}$ is modeled as  
$$
\pi_{iml}=\frac{\exp(w_{iml})}{\sum_{l'=1}^L\exp(w_{iml'})},  \ \ \ \ \ (w_{1ml},\ldots,w_{nml})\sim {\cal N}(0, \tau^{-1} Q(\psi)), 
$$ 
where $Q(\psi)$ is a correlation matrix with unknown parameter $\psi$.
\end{description}

The prediction MSE values are averaged over 20 replications, and the results are shown in Table~\ref{tab:sim}.
Overall, the proposed sIBP provides better prediction accuracy than the other models. 
In particular, IBP does not consider spatial dependence and cannot predict unobserved locations while accounting for local spatial features, while sIBP achieves more accurate predictions owing to the latent Gaussian process. 
On the other hand, while MnGP can model spatial trends of multinomial probabilities for accurate predictions, it struggles to capture discontinuous structures in spatial trends, leading to sIBP achieving higher accuracy than MnGP.

%  Table 
\begin{table}
\caption{The mean squared errors (MSE) of prediction in test samples, produced by sIBP, IBP, XGBoost and MnGP, averaged over 20 Monte Carlo replications.  
\label{tab:sim}}
\vspace{1.5ex}
\centering
\begin{tabular}{cccccccc}
\hline
Scenario & sIBP & IBP & XGBoost & MnGP \\
\hline
(I) & \textbf{2.82} & 4.38 & 5.51 & 3.42 \\
(II) &  \textbf{2.04} & 5.62 & 6.67 & 2.39 \\
(III) & \textbf{0.59} & 2.04 & 0.88 & 0.77 \\
\hline
\end{tabular}
\end{table}

%----------------------------------------------------------%
%                  Data analysis                           %
%----------------------------------------------------------%
\section{Applications}
\label{sec:app}

\subsection{Spatial factor analysis of dialectal data}\label{sec:dialect}

We here demonstrate the spatial factor models for multivariate multinomial data
described in Section~\ref{sec:multinomial-model}, using the real dataset
regarding the geographical distributions of dialects observed in Japan. 
The observed data is an $N\times M$ matrix $X$ whose $(n,m)$-element $x_{nm}$ \ ($n=1,\ldots,N$ and $m=1,\ldots,M$) takes one of $\{1,\ldots,c_m\}$, corresponding to the word form used for representing $m$'th
concept at the $n$'th location.
Note that $c_m$ is the number of different word forms used over the $N$ locations.
Analyzing this type of dataset is substantial in dialectology, and several
statistical approaches exist such as network models
\citep{lee2011bayesian,saitou2017language} and latent Dirichlet allocation
\citep{syrjanen2016applying,cathcart2020probabilistic}.
Further, \cite{murawaki2020latent} developed a geographical factor model
similar to the model in Section~\ref{sec:multinomial-model}, with different
probabilistic models for latent binary variables and the number of factors
being fixed.
Here, we apply the spatial factor model with sIBP to dialectal data in Japan to
extract latent structures, including the number of factors.

%\subsubsection{Dialects in Japanese}

\begin{figure}[t]
 \centering
 {\small \tabcolsep=4pt
 \begin{tabular}{ccc}
  \includegraphics[width=0.28\linewidth]{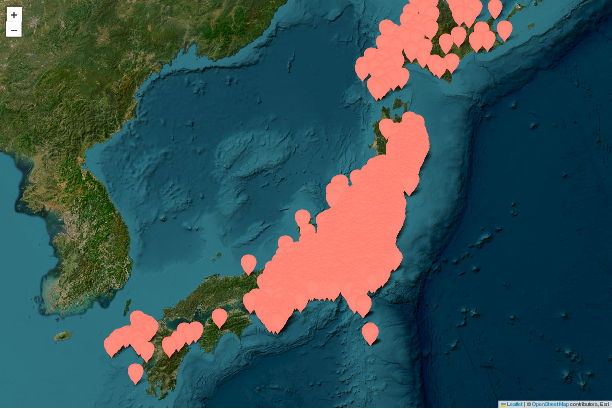} &
  \includegraphics[width=0.28\linewidth]{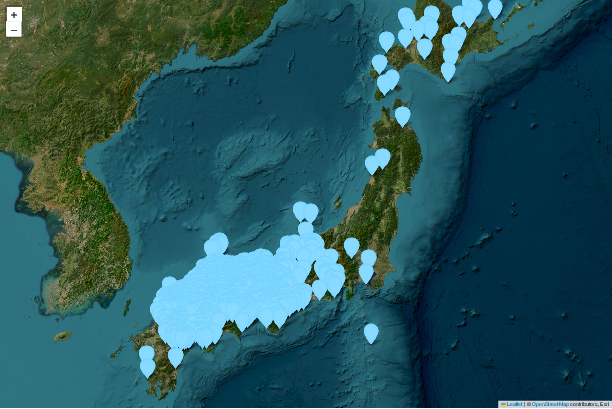} &
  \includegraphics[width=0.28\linewidth]{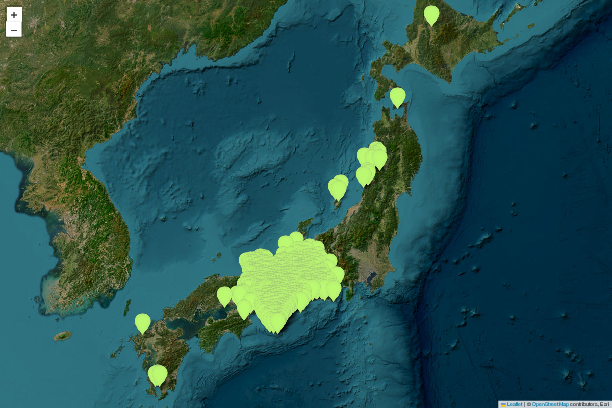} \\
  Factor 1 & Factor 2 & Factor 3 \\[1ex]
  \includegraphics[width=0.28\linewidth]{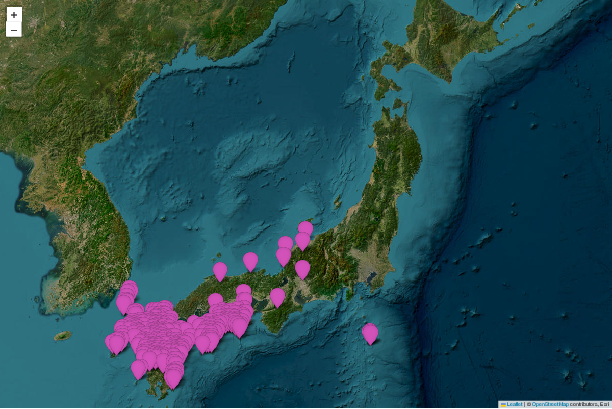} &
  \includegraphics[width=0.28\linewidth]{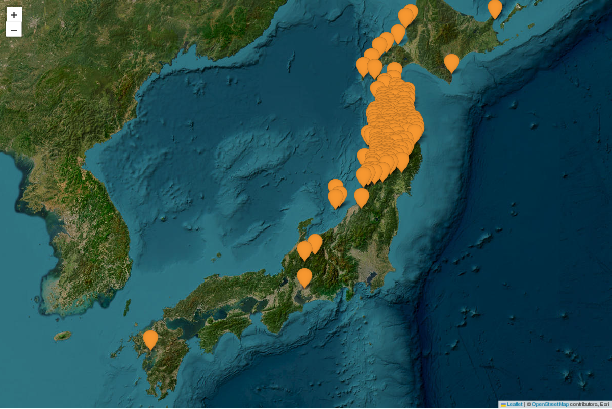} &
  \includegraphics[width=0.28\linewidth]{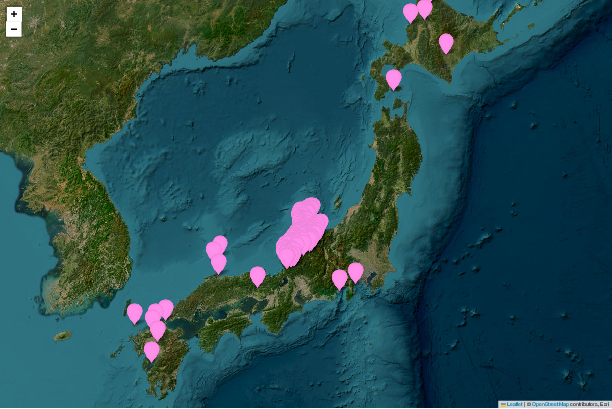} \\
  Factor 4 & Factor 5 & Factor 6 \\[1ex]
  \includegraphics[width=0.28\linewidth]{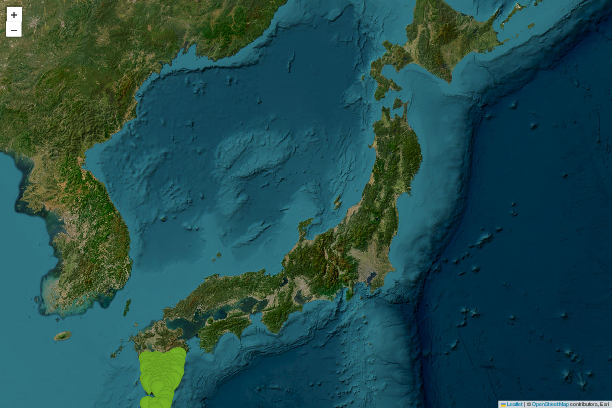} &
  \includegraphics[width=0.28\linewidth]{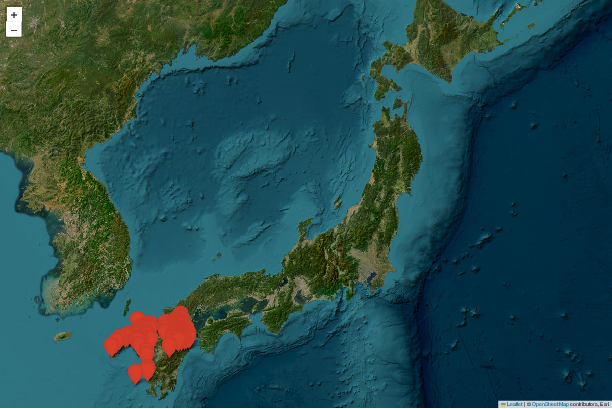} &
  \includegraphics[width=0.28\linewidth]{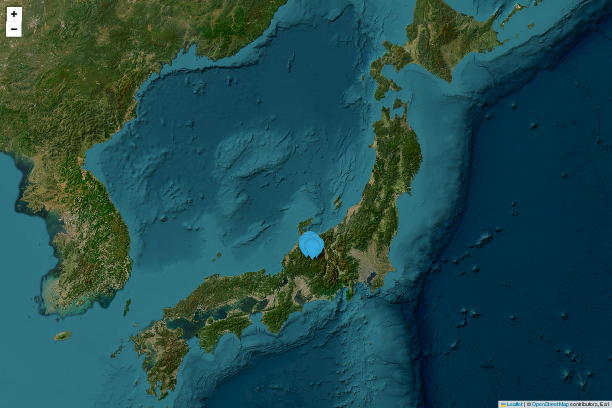} \\
  Factor 7 & Factor 8 & Factor 9
 \end{tabular}
 }
 \caption{Geographical distribution of inferred factors by sIBP for Japanese dialects.}
 \label{fig:factor-japan}
 \vspace{1em}
 \includegraphics[width=0.9\linewidth]{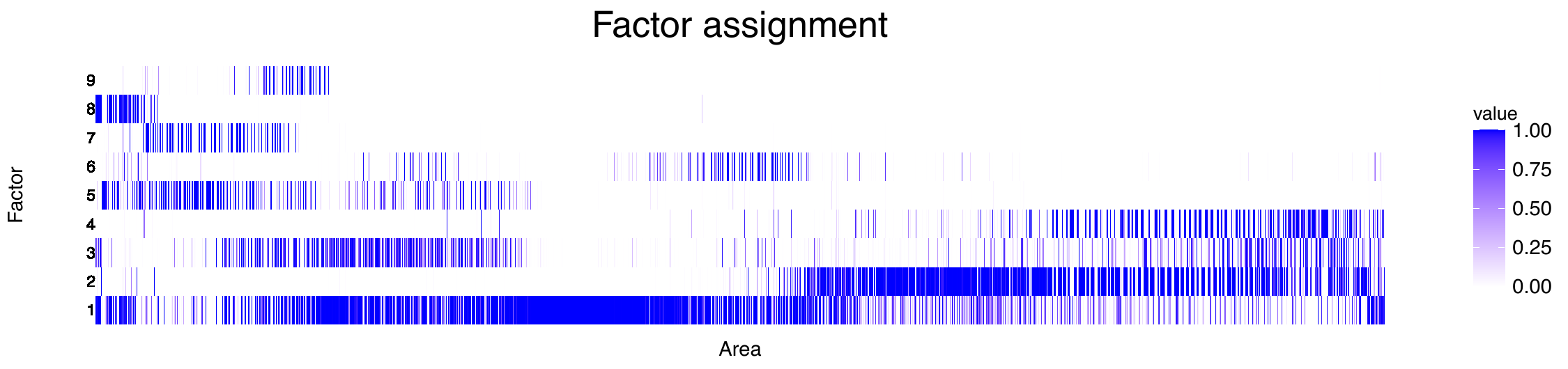}
 \caption{Posterior probability of factors for each of $2,317$ locations.
 Locations are ordered from north (left) to south (right).
 Since nothern Hokkaido island (leftmost) is a place of immigrants,
 it has a mixture of factors from their native places.
 }
 \label{fig:factor-japan-dist}
\end{figure}

For the dataset, we employed a subset of the dialectal data in Japan, called 
Linguistic Atlas of Japan (LAJ)~\citep{lajdb}\footnote{
This dataset is publicly available from \url{https://www.lajdb.org/}.
}, which has $N\!=\!2317$ locations and $M\!=\!27$ features. 
We excluded southern Ryukyu islands because they are quite distant from the
main islands of Japan and have distinct dialects.
For simplicity, we unified original word forms with the first letter to reduce
large heterogeneity: for example, word forms ``atama'', ``adama'', ``anma''
for a concept ``\begin{CJK}{UTF8}{min}あたま\end{CJK}'' (head) are unified as ``\begin{CJK}{UTF8}{min}あたま\end{CJK}/a/''.
As a result, $c_m$ ranges from $2$ to $14$, with the average value being
$8.1$. \footnote{
For the original LAJ dataset, number of different word forms $c_m$ ranges
from $5$ to $470$ with the average $103.1$.
}

While there are approximately $5\%$ missing entries in the observed matrix $X$,
it can be safely handled by our MCMC algorithm by imputing these missing values
via data augmentation in each MCMC iteration.  Since the number of locations is
large, we apply the scalable version of the sIBP with $K\!=\!10$ 
(the largest number of factors) 
and $m\!=\!10$ nearest-neighbor Gaussian process with exponential kernel function, 
as described in Section~\ref{sec:scalable}.
We generated $10,000$ posterior samples after discarding the first $5,000$ samples
as burn-in.

Based on the posterior samples, we computed the posterior probability being
$z_{ik}\!=\!1$ and found that the 10th factor includes only two locations, 
so we regard the number of latent factors as $9$.
For each location $i\!=1,\ldots,2317$ and factor $k\!=\!1,\ldots,9$, 
we computed the posterior probability being 
$z_{ik}\!=\!1$ and show locations having a probability greater than $0.5$ in
Figure~\ref{fig:factor-japan}, and 
factors possessed by each location in Figure~\ref{fig:factor-japan-dist}.
It reveals that the geographical distribution of the detected factors varies
significantly from one another, and the number of locations constituting each
geographical factor decreases monotonically as the index of factors
(vertical axis in Figure~\ref{fig:factor-japan-dist}),
which seems desirable for interpretability. 
Such property would come from the definition of the IBP that gives
monotonically decreasing prior probability over the factors.

\begin{figure}[t]
 \centering
 {\footnotesize \tabcolsep=9pt
 \begin{tabular}{cc}
  \includegraphics[height=6cm,trim=65 0 100 0,clip]{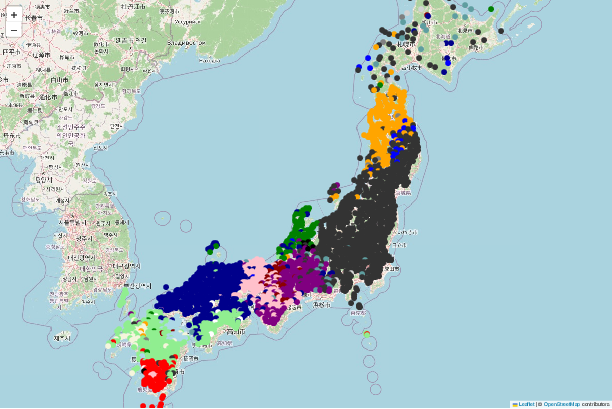}
  &
  \includegraphics[height=6cm,trim=70 137 12 35,clip]{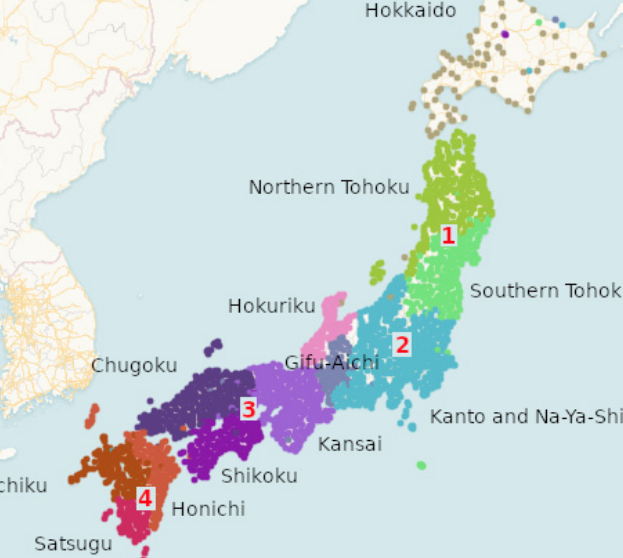}
  \\[0.1em]
  \cellbox{0.25\textwidth}{(a) sIBP-induced clusters\\[2pt] 
           \centering (this study)} & 
  \cellbox{0.25\textwidth}{\centering (b) HAC (replicated from\\[2pt] 
           \cite{heeringa23laj})}
 \end{tabular}
 }
 \vspace{-1ex}
 \caption{Comparison of different areal clusterings.
 The number of clusters are $15$ and $13$.
 }
 \label{figure:cluster-comparison}
\end{figure}

%\paragraph{Qualitative evaluation}
Based on the $9$-bit encoding of each location computed from 
Figure~\ref{fig:factor-japan} and \ref{fig:factor-japan-dist},
we also have a clustering of locations shown in 
Figure~\ref{figure:cluster-comparison}(a),
which has $15$ clusters with size $\geq 10$.
Induced clustering of locations basically similar to the previous work of
\citet{heeringa23laj} that uses a simple hierarchical agglomerative clustering (HAC)
shown in Figure~\ref{figure:cluster-comparison}(b), which happens to have a similar
number of $13$ clusters excluding Ryukyu islands.

These two areal clusterings align with each other such as a clear boundary between
eastern and western dialects, Kansai (Kyoto-Osaka area) dialects,
Hokuriku and Kyushu dialects.
However, there are also essential differences not found in previous HAC approaches.
For example, southern part of Shikoku island shares the same cluster as a part of
Kyushu island, which is intuitive considering maritime transportation prevalent
in older ages.
Southern peninsula of Kansai area belongs to the same cluster as Nagoya area,
which is much larger than that of HAC and matches well with current understanding of
Japanese dialects,
might also be due to maritime transportation; 
southern Tohoku area is unified with Kanto (Tokyo area) for our sIBP analysis.
Note that contrary to ordinary post-hoc areal clustering like
Figure~\ref{figure:cluster-comparison}(b),
our sIBP clustering in Figure~\ref{figure:cluster-comparison}(a)
can be further decomposed into different latent factors shown in Figure~\ref{fig:factor-japan}.
While southern Shikoku island has similarity to Kyushu as described above
(Factor 4), Kyushu has its own latent factor (Factor 8).
Our sIBP factorization is able to simultaneously describe commonality and
idiosyncrasy between the areas, which is a distinctive characteristics of our model.

\begin{figure}[t]
 \medskip
 \centering
 {\footnotesize \tabcolsep=9pt
 \begin{tabular}{ccc}
  \includegraphics[width=0.27\textwidth]{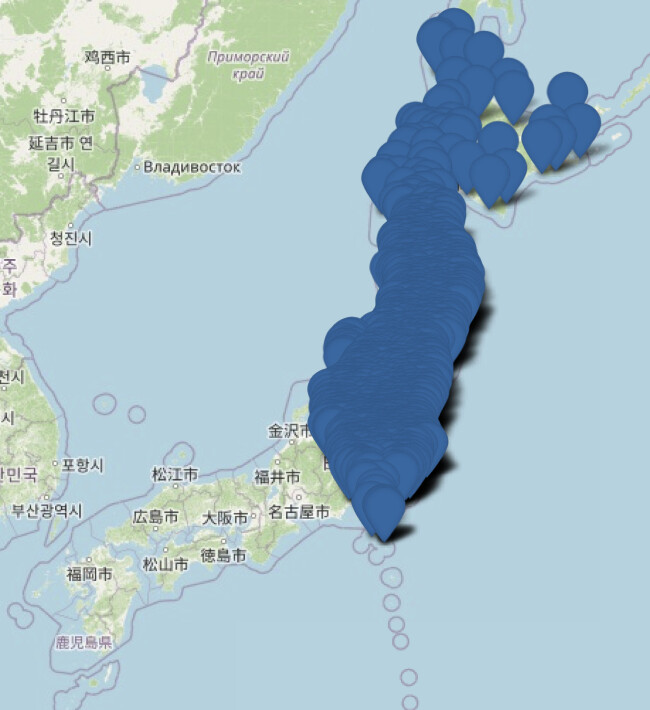} &
  \includegraphics[width=0.27\textwidth]{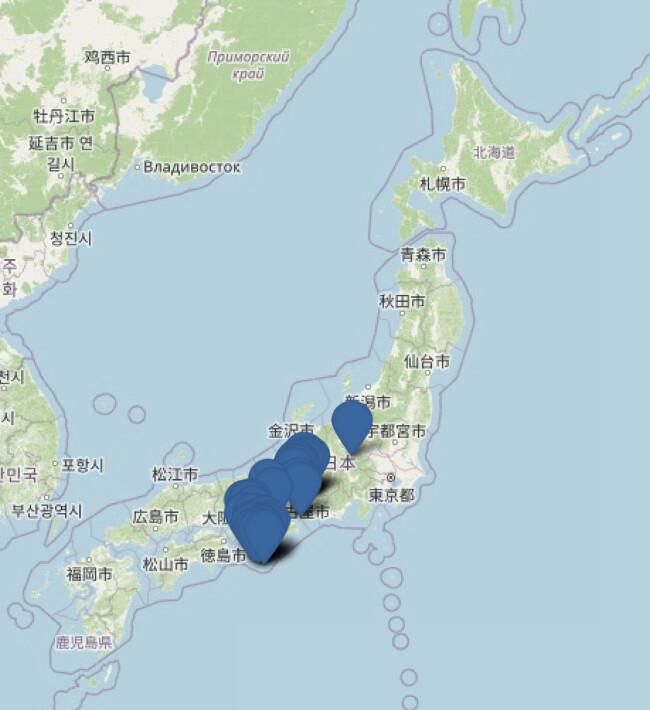} &
  \includegraphics[width=0.27\textwidth]{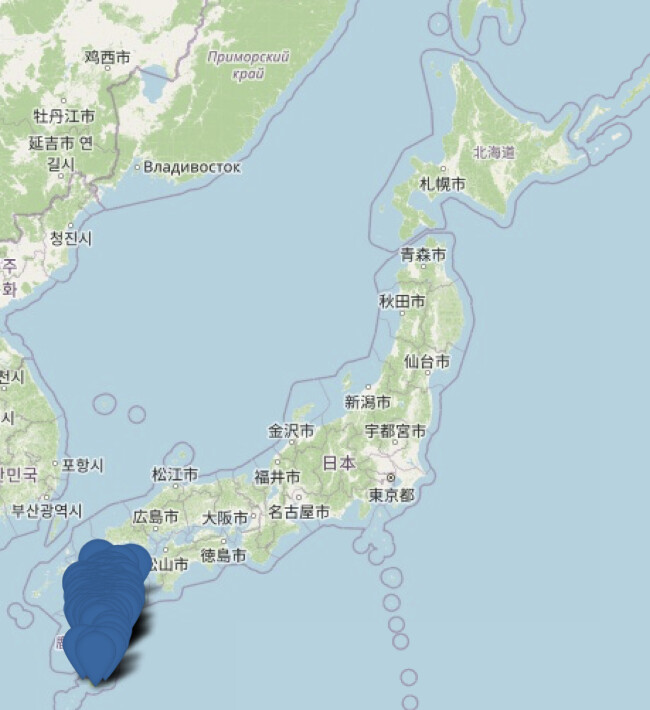} \\[1ex]
  (a) \begin{CJK}{UTF8}{min}しあさって\end{CJK}/y/ &
  (b) \begin{CJK}{UTF8}{min}あたま\end{CJK}/s/ &
  (c) \begin{CJK}{UTF8}{min}あたま\end{CJK}/b/ \\
  (for Factor 1) & (for Factor 3) & (for Factor 7)
 \end{tabular}
 }
 \caption{Distribution of dialectal forms in LAJ.}
 \label{figure:word-map}
\end{figure}

\begin{table}[t]
 \caption{Words with the highest PMI$\mbox{}^{{\kern1pt}2}$ values 
 for each factor in LAJ dataset.
 Geographical distributions of \uwave{underlined} forms are shown in 
 Figure~\ref{figure:word-map}.
 }
 \label{table:factor-pmi-words}
 \vspace{-1ex}
 \begin{center}
 \includegraphics[width=\textwidth]{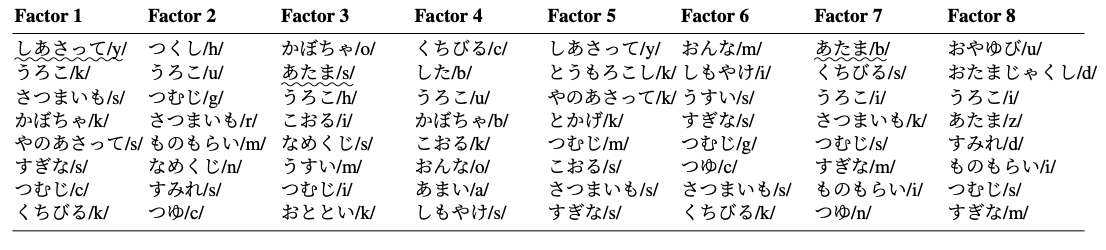} 
 \end{center}
\end{table}

According to sIBP, each latent factor $k$ has a distribution over word forms (dialects)
$1,\cdots, c_m$ for each concept $m$.
Table~\ref{table:factor-pmi-words} shows dialects that are most associated with
each factor, some of which are also displayed in Figure~\ref{figure:word-map}.
Prominent dialects are actually distributed along each factor, 
validating our model for spatial heterogeneity of dialects.

Since some dialects have equally high probability for all the factors,
to obtain Table~\ref{table:factor-pmi-words} we computed a statistic
that is proportional to PMI${}^2$~\citep{bouma09npmi}, namely
\begin{equation}
 \log \frac{p(w|c)}{\sqrt{p(w)}} = \log p(w|c) - \frac{1}{2}\log p(w)
 \label{eqn:pmi2}
\end{equation}
for each word $w$ and each factor $c$.
Equation~\eqref{eqn:pmi2} is a modified version of Pointwise Mutual Information (PMI)
$\log p(w|c)/p(w)$ \citep{church89association},
but avoids overfitting to rare dialects whose $p(w)$ is very small.%
\footnote{
Na\"ively using PMI yields very rare words being associated with each factor.
This statistic is also proportional to the weighting used in Canonical Correlation
Analysis (CCA) that is proved to work very well for linguistic contingency tables
\citep{stratos15cca}.
}

%\paragraph{Quantitative evaluation}
We also conducted quantitative validation of sIBP.
If the clusters adequately describe dialects, dialects in each cluster should be
similar: in other words, dialects in the same cluster $c$ should have been generated
from the same distribution $\theta_c$.
Based on this intuition, we assume a simple Bayesian generative model for validation:
for each concept $m$,
(a) each cluster has a latent multinomial distribution $\theta_c$ 
drawn from a Dirichlet distribution ${\rm Dir}(\bm{\alpha})$, and
(b) each dialectal form $x$ within this cluster
is drawn from $\theta_c$.
Since we do not know $\theta_c$ beforehand, we can integrate it out:
{\abovedisplayskip=8pt \belowdisplayskip=8pt
\begin{align}
 p(x|c)
 &= \int p(x|\theta_c)\,p(\theta_c)d\theta_c 
  = \frac{\Gamma(\sum_{k=1}^{c_m} \alpha_k)}{\Gamma(\sum_{k=1}^{c_m} \alpha_k \!+\! n_k)}
    \prod_{k=1}^{c_m} \frac{\Gamma(\alpha_k \!+\! n_k)}{\Gamma(\alpha_k)}
    \label{eqn:polya-likelihood}
\end{align}
}
which is known as P\'olya distribution~\citep{minka00dirichlet,murphy22pml}.
Here, $c_m$ is the number of different word forms for the concept $m$ and $n_k$ is the number of locations that employ $k$'th word form in this cluster.
Using a uniform Dirichlet distribution $\alpha_1=\ldots=\alpha_{c_m}=1$, we can quantitatively compare different clusterings using the marginal likelihood of \eqref{eqn:polya-likelihood} of the observations with minimum assumptions.

The result is shown in Table~\ref{table:word-likelihood}.
Compared to ordinary hierarchical agglomerative clustering (HAC) employed in 
previous studies \citep{heeringa23laj},
our sIBP achieved much higher marginal likelihood than HAC,
while HAC itself has higher likelihood than random clusterings (both the number of
clusters $C\!=\!15$ and $50$),
showing that the marginal likelihood actually identifies the quality of clusterings.

\begin{table}[b]
 \caption{
 Log likelihood and the effective number of clusters (clusters with size $\geq 10$)
 for each model. sIBP achieved the best marginalized likelihood over hierarchical
 agglomerative clustering (\cite{heeringa23laj}) and random cluster assignments.
 }
 \label{table:word-likelihood}
 \vspace{0.6em}
 \centering
 \begin{tabular}{r|cccc}
  \tc{\multirow{2}{*}{Model}}\vrule
  & \tc{sIBP} & \tc{HAC} & \tc{Random} & \tc{Random} \\
  &
  ($\theta\!=\!0.5$) &
  ($C\!=\!50$) &
  ($C\!=\!15$) &
  ($C\!=\!50$) 
  \\ \hline
  Effective \# of clusters &
  $15$ & $13$ & $15$ & $50$ \\
  Log marginalized likelihood ($\uparrow$) &
  $\mathbf{-26567.4}$ & $-37970.4$ & $-53124.2$ & $-57984.2$
  \\ \hline
 \end{tabular}
\end{table}

\subsection{Factor analysis of geographical tree distribution}\label{sec:tree}

We next apply the sIBP model to tree census data in Japan reported in \citep{ishihara2011forest}.
The data is the number of trees (with girth at breast height of 15 cm or more) for each species, collected at 42 locations (typically one hectare in each location), covering subarctic to subtropical climate zones and the four major forest types in Japan.
By eliminating one outlying location and selecting tree species observed in at least 10 locations, the dataset we apply consists of $n=41$ locations and $M=45$ species, 
which is spatially sparser than dialectal data in the previous subsection.

We fit the multivariate count factor model described in Section~\ref{sec:count-model} with sIBP and IBP. 
For sIBP, the standard Gaussian process with the exponential kernel is used, and the maximum number of factors is set to $K=7$ for both models.
By generating 5,000 posterior samples after discarding the first 5,000 samples as burn-in, we calculated the posterior probability of the possession of factors. 
We detected four non-empty factors by sIBP and three non-empty factors by IBP.

\begin{figure}[htbp!]
\centering
\includegraphics[width=\linewidth]{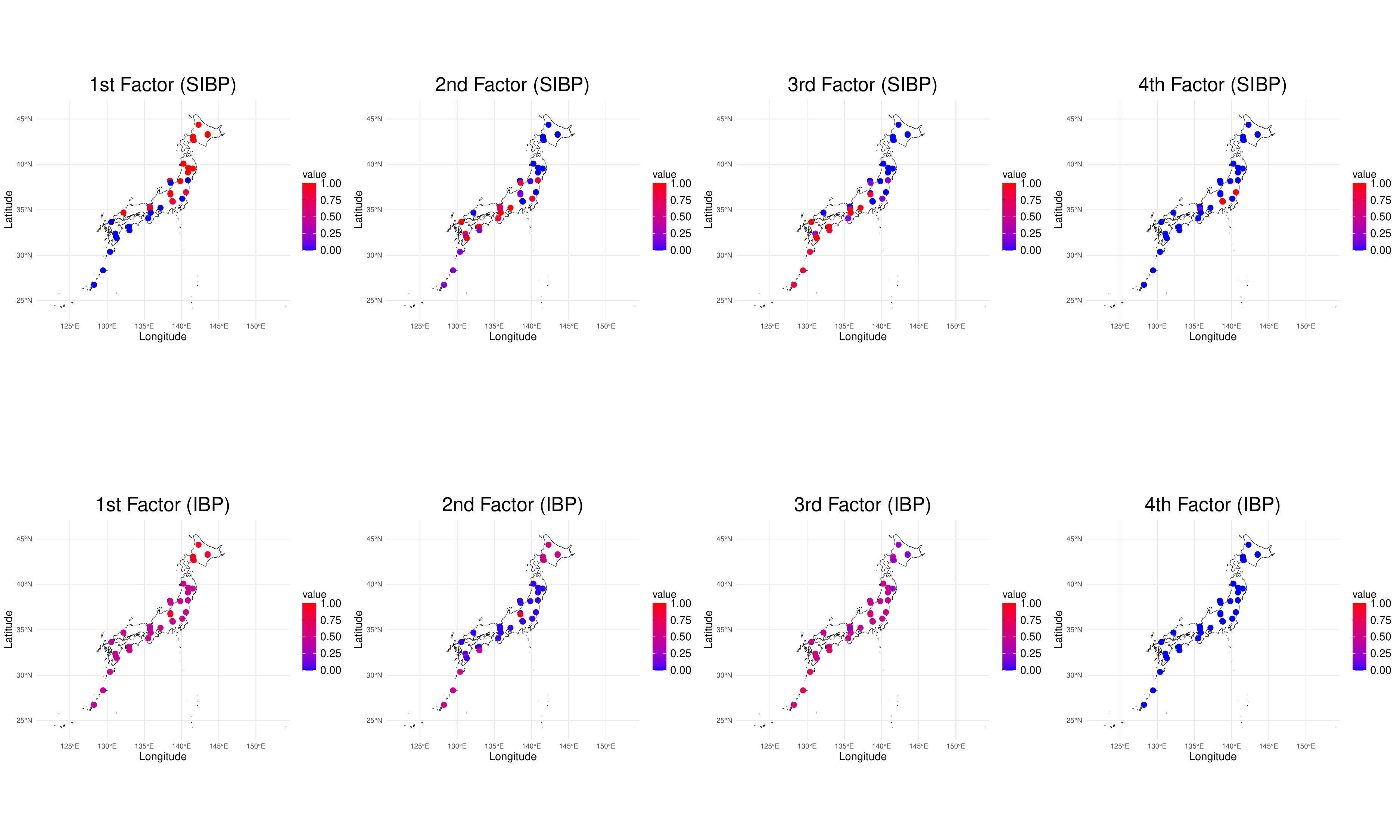}
\caption{Geographical distribution of posterior probabilities of the presence of four factors, obtained by sIBP (upper) and IBP (lower). } 
\label{fig:factor-tree}
\end{figure}

\begin{figure}[htbp!]
 \centering
 {\small
 \begin{tabular}{cc}
         & Intensity of Species \\[1ex]
  Factor & \includegraphics[width=0.7\linewidth,align=c,trim=40 20 0 20,clip]{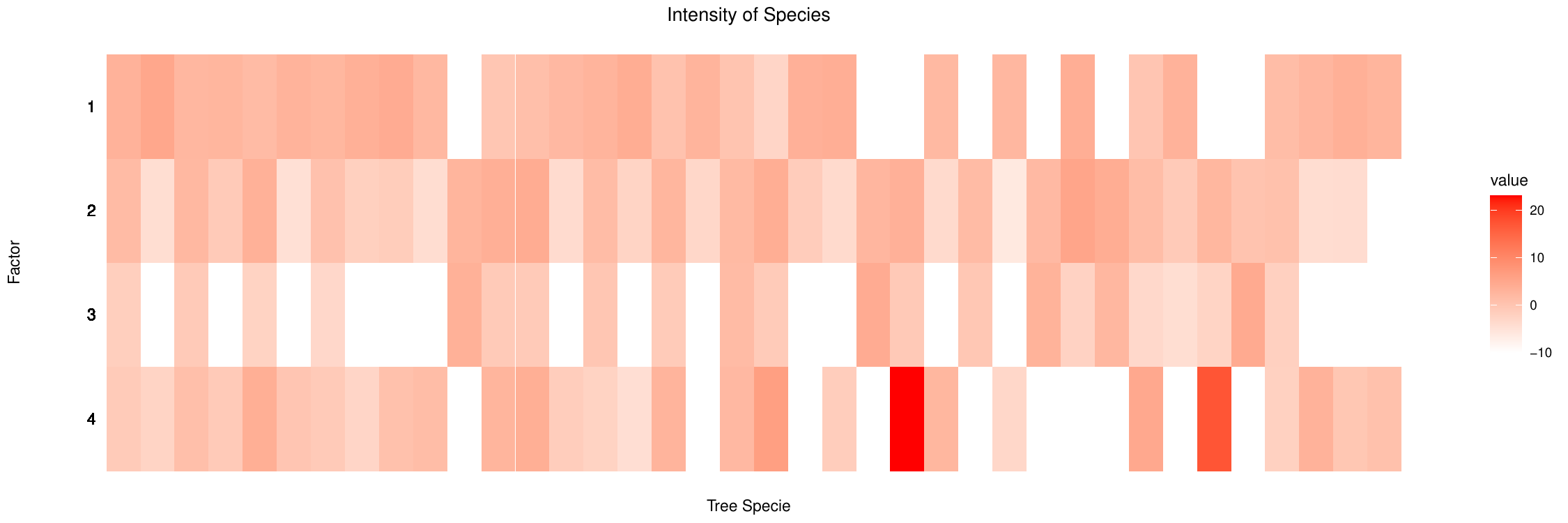} \\
         & Tree Species\rule{0pt}{1.5em}
 \end{tabular}
 }
 \caption{Posterior means of log-intensity of each species in four factors, obtained by sIBP.} 
 \label{fig:factor-tree-dist}
\end{figure}

Figure~\ref{fig:factor-tree} shows the geographical distribution of the posterior probabilities of factor presence. 
The factors obtained by IBP are difficult to interpret, since the first and third factors are assigned high posterior probabilities at almost all locations, suggesting that IBP mainly produces broad factors that do not clearly distinguish regional vegetation patterns. 
In contrast, sIBP detects four spatially structured factors: one broadly shared component and several additional components capturing regional deviations. 
The factor-specific log-intensities in Figure~\ref{fig:factor-tree-dist} further show that these factors have distinct species profiles, indicating that the detected factors can be interpreted as latent vegetation components characterized jointly by their spatial support and species composition. 
The deviance information criterion \citep{spiegelhalter2002bayesian} is 5,769 for sIBP and 6,113 for IBP, further supporting the superior fit of sIBP for this dataset.

\begin{figure}[htbp!]
\centering
\includegraphics[width=\linewidth]{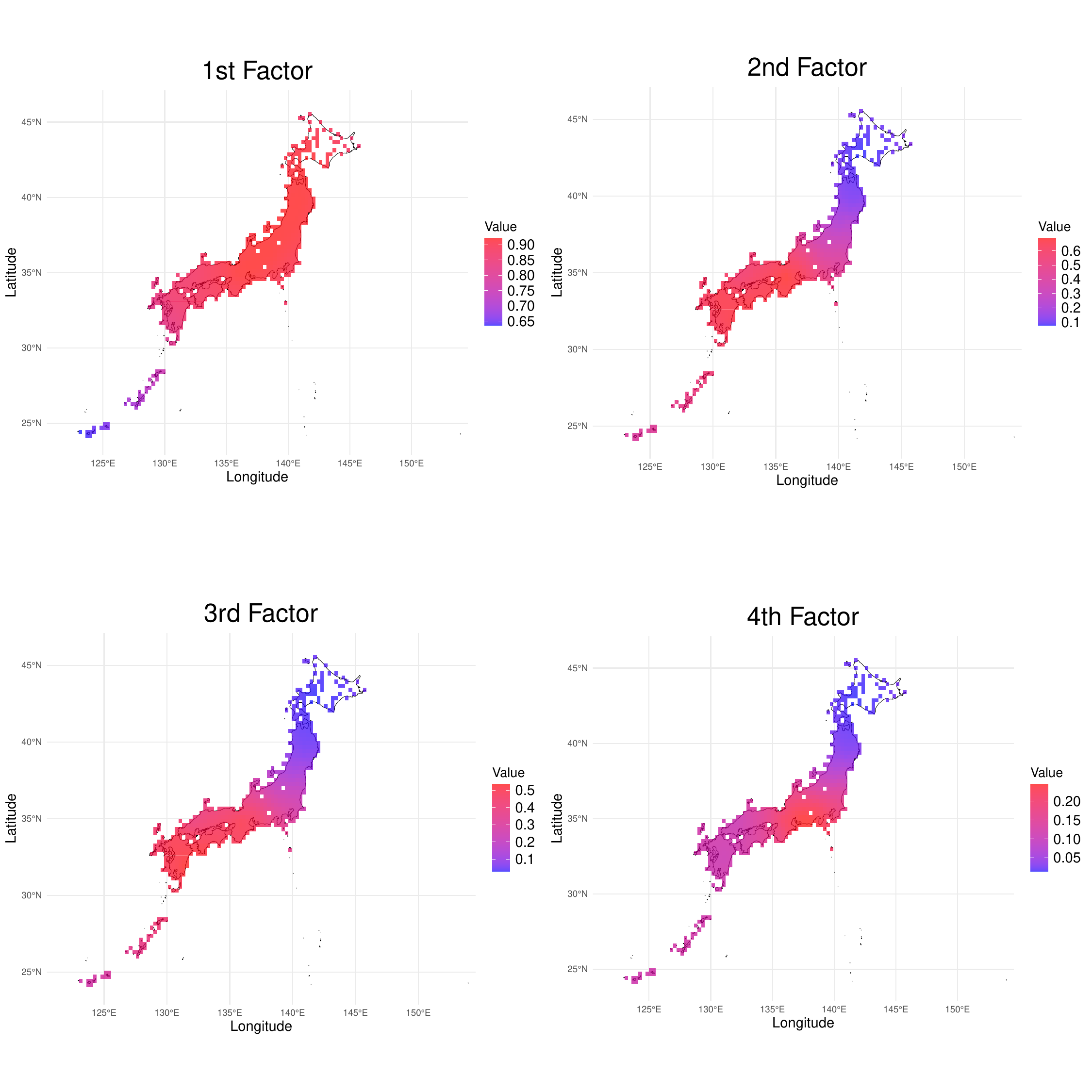}
\caption{Predicted probability of the presence of the four factors, obtained by sIBP.} 
\label{fig:factor-tree-smooth}
\end{figure}

Finally, leveraging the latent Gaussian process, we can predict the probability of factor presence at arbitrary locations, which cannot be done without process-based spatial modeling. 
Figure~\ref{fig:factor-tree-smooth} presents the predicted probability surfaces of the four factors. 
The first factor has high predicted probability over a wide range of mainland Japan and can be viewed as a broadly shared vegetation component. 
The other factors exhibit clearer spatial gradients, with relatively high probabilities in specific regions and lower probabilities elsewhere. 
These smooth probability maps provide an interpretable summary of how latent vegetation components vary continuously over space. 
Moreover, by combining these predicted factor probabilities with the factor-specific log-intensities in Figure~\ref{fig:factor-tree-dist}, the model can be used to predict tree species composition at unobserved locations.

%--------------------------------------------%
%          Concluding remarks                %
%--------------------------------------------%
\section{Concluding Remarks}
We proposed sIBP, a generalization of the standard IBP, by incorporating spatial information through a latent Gaussian process. 
While we focus on two-dimensional spatial information, it can be immediately extended to general covariates; that is, two subjects with similar covariates tend to share the same factors. 
Such generalization would be useful in general latent feature (factor) models such as item response theory models \citep{li2023sparse}.
Further, the proposed sIBP is worth developing its spatio-temporal version or fast computation algorithm for sIBP using, for example, variational approximation as done in \cite{doshi2009variational} for the standard IBP. 
A detailed investigation would extend the scope of this paper, so that is left to future studies. 

%-----------------------------------------------%
%             Acknowledgement                   %
%-----------------------------------------------%
\section*{Acknowledgement}
This work is supported by the Japan Society for the Promotion of Science (JSPS KAKENHI) grant numbers 
18KK0012, 20H00080, and 21H00699.

%-----------------------------------------------%
%                Reference                      %
%-----------------------------------------------%
\vspace{0.5cm}
\bibliographystyle{chicago}
\bibliography{ref}

\newpage
%-------------------------------------------%
%              Supplement                   %
%-------------------------------------------%
\setcounter{equation}{0}
\setcounter{section}{0}
\setcounter{table}{0}
\setcounter{figure}{0}
\setcounter{page}{1}
\renewcommand{\thesection}{S\arabic{section}}
\renewcommand{\theequation}{S\arabic{equation}}
\renewcommand{\thetable}{S\arabic{table}}
\renewcommand{\thefigure}{S\arabic{figure}}

\vspace{1cm}
\begin{center}
{\LARGE \textbf{Supplementary Material for ``Spatially Dependent Indian Buffet Processes''}}
\end{center}

\section{Proof of Proposition~1}
It is sufficient to show that the finite-dimensional distributions
induced by the construction are projectively consistent. 
Fix arbitrary
locations $s_1,\ldots,s_n\in\mathcal{S}$. 
For each $k$, the latent process $u_k(\cdot)$ is a Gaussian process, and hence $\{u_k(s_1),\ldots,u_k(s_n)\}$, has a well-defined multivariate normal distribution. 
Moreover, the finite-dimensional distributions of $\{u_k(\cdot)\}$ are consistent under permutations and marginalization over locations.

For a finite truncation level $L$, define $b_k(s_i)=\prod_{j=1}^k \sigma\{u_j(s_i)\}, \ i=1,\ldots,n, \  k=1,\ldots,L$.
Given the latent variables $\{u_j(s_i):i=1,\ldots,n,\ j=1,\ldots,L\}$,
the binary feature indicators are generated according to the conditional
Bernoulli distributions specified in (\ref{Sp-IBP-process}). Thus, for
any $a_{ik}\in\{0,1\}$, it holds that 
$$
\begin{aligned}
&\mathbb{P}\Big(z_k(s_i)=a_{ik},\ i=1,\ldots,n,\ k=1,\ldots,L
\mid \{u_j(s_i)\}\Big)  
=\prod_{k=1}^L\prod_{i=1}^n
b_k(s_i)^{a_{ik}}\{1-b_k(s_i)\}^{1-a_{ik}} .
\end{aligned}
$$
Integrating this conditional distribution with respect to the joint
Gaussian distribution of the latent variables gives a valid joint
distribution for $\{z_k(s_i):i=1,\ldots,n,\ k=1,\ldots,L\}$.
These distributions are consistent as $L$ varies. 
Indeed, marginalizing over the indicators corresponding to feature $L+1$ gives one, and $b_k(s_i)$ for $k\leq L$ does not depend on $u_{L+1}(\cdot)$. Therefore, after integrating out $u_{L+1}(s_1),\ldots,u_{L+1}(s_n)$, the resulting distribution coincides with the distribution for the first $L$ features.
Hence, by the Kolmogorov's extension theorem applied to the feature index,
there exists a joint distribution for $\{Z(s_1),\ldots,Z(s_n)\}, \ 
Z(s_i)=\{z_1(s_i),z_2(s_i),\ldots\}$. 

It remains to check consistency with respect to the spatial locations.
Consider an additional location $s_{n+1}$. 
By the projective consistency
of Gaussian process finite-dimensional distributions, marginalizing the
joint distribution of $\{u_k(s_1),\ldots,u_k(s_n),u_k(s_{n+1})\}$ over $u_k(s_{n+1})$ gives the joint distribution of $\{u_k(s_1),\ldots,u_k(s_n)\}$.
Moreover, conditional on the latent variables, summing over
$z_k(s_{n+1})\in\{0,1\}$ gives one for each $k$. 
Therefore, marginalizing
over $Z(s_{n+1})$ gives the same finite-dimensional distribution as that
for $\{Z(s_1),\ldots,Z(s_n)\}$. 
The same argument also shows invariance under permutations of the locations.

Thus, the finite-dimensional distributions induced by the
infinite-feature version of (\ref{Sp-IBP-process}) are projectively
consistent over $\mathcal{S}$. Since $\{0,1\}^{\mathbb{N}}$ equipped with
the product sigma-field is a standard Borel space, Kolmogorov's extension
theorem guarantees the existence of a probability measure on
$(\{0,1\}^{\mathbb{N}})^{\mathcal{S}}$ whose finite-dimensional marginals
are those specified above. Consequently, the collection
$\{Z(s):s\in\mathcal{S}\}$ is well-defined as a spatial stochastic
process on $\mathcal{S}$.

\section{Proof of Proposition~2}
Since $u_{i1},\ldots,u_{ik}$ are independent and their marginal distributions are ${\cal N}(\mu, \tau^{-1})$, it follows that 
$$
\mathbb{P}(z_{ik}=1)=\mathbb{E}[z_{ik}]=\mathbb{E}\left[\prod_{j=1}^k\sigma(u_{ij})\right]=\delta_1^k.
$$
Since $\sigma(\cdot)$ is a logistic function, $\delta_1\in (0,1)$, so that we have $\mathbb{P}(z_{ik}=1)\to 0$ as $k\to\infty$. 
Using the above results, we also have 
$$
\mathbb{E}[c_i]=\sum_{j=1}^k \mathbb{E}[z_{ij}]=\sum_{j=1}^k\delta_1^j=\delta_1\cdot \frac{1-\delta_1^k}{1-\delta_1}\to  \frac{\delta_1}{1-\delta_1}>0
$$
as $k\to\infty$.
Furthermore, letting $u_i=(u_{i1},\ldots,u_{iK})$, we have 
\begin{align*}
{\rm Var}\left(c_i\right)
&={\rm Var}\big(\mathbb{E}[c_i|u_i]\big)
+ \mathbb{E}\big[{\rm Var}(c_i|u_i)\big] \\
&={\rm Var}\left(\sum_{k=1}^Kb_{ik}\right)
+ \sum_{k=1}^K \mathbb{E}\left[b_{ik}(1-b_{ik})\right]\\
&=\mathbb{E}\left[\Big(\sum_{k=1}^Kb_{ik}\Big)^2\right] - \left\{\sum_{k=1}^K \mathbb{E}[b_{ik}]\right\}^2 + \sum_{k=1}^K \mathbb{E}[b_{ik}] - \sum_{k=1}^K \mathbb{E}[b_{ik}^2]\\
&=2\sum_{j=1}^K \sum_{k=1}^{j-1} \mathbb{E}[b_{ij}b_{ik}]  - \left\{\sum_{k=1}^K \mathbb{E}[b_{ik}]\right\}^2 + \sum_{k=1}^K \mathbb{E}[b_{ik}] 
\end{align*}
where we used conditional independence of $z_{i1},\ldots,z_{iK}$ given $u_i$.
Note that $\mathbb{E}[b_{ik}]=C_1^k$ and $\mathbb{E}[b_{ij}b_{ik}] =\delta_2^k \delta_1^{j-k}$ for $k<j$,
Then, it follows that 
\begin{align*}
\sum_{j=1}^K \sum_{k=1}^{j-1} \mathbb{E}[b_{ij}b_{ik}] 
&=\sum_{j=1}^K  \delta_1^j\sum_{k=1}^{j-1}  \left(\frac{\delta_2}{\delta_1}\right)^k \\
&=\frac{\delta_1\delta_2}{\delta_1-\delta_2}\left(\frac{1-\delta_1^K}{1-\delta_1} - \frac{1-\delta_2^K}{1-\delta_2}\right)
\to \frac{\delta_1\delta_2}{(1-\delta_1)(1-\delta_2)}
\end{align*}
as $K\to\infty$. 
Hence, we have 
\begin{align*}
\lim_{K\to\infty} {\rm Var}(c_i)&=\frac{2\delta_1\delta_2}{(1-\delta_1)(1-\delta_2)} - \left(\frac{\delta_1}{1-\delta_1}\right)^2+\frac{\delta_1}{1-\delta_1}\\
&=\frac{\delta_1}{1-\delta_1} \left(1+\frac{2\delta_2}{1-\delta_2}-\frac{\delta_1}{1-\delta_1}\right),
\end{align*}
which completes the proof. 

\section{Proof of Proposition~3}
First, we note that 
\begin{align*}
\mathbb{P}(z_{ik}=1, z_{i'k}=1)=\mathbb{E}[b_{ik}b_{i'k}]=\prod_{j=1}^k \mathbb{E}[\sigma(u_{ij})\sigma(u_{i'j})].
\end{align*}
Without loss of generality, we assume that $\mu=0$ and $\tau=1$ in what follows.  
Since $(u_{ij}, u_{i'j})\sim {\cal N}((0,0), R)$ with $R_{11}=R_{22}=1$, $R_{12}=R_{21}=\rho_{ii'}$ and $\rho_{ii'}=\rho(\|s_i-s_{i'}\|; \psi)$, we have 
\begin{align*}
D_{\mu,\tau}(\rho_{ii'})\equiv \mathbb{E}[\sigma(u_{ij})\sigma(u_{i'j})]=
\iint \sigma(x)\sigma(y)\phi_2((x,y); (0,0), R)dxdy,
\end{align*}
where $\Sigma_{\tau}(\rho_{ii'})$ is the $2\times 2$ covariance matrix, and this integral does not depend on $j$. 
We note that 
\begin{align*}
\frac{\partial}{\partial \rho} \phi_2((x,y); (0,0), R)
&=\frac{\phi_2((x,y); (0,0), R)}{1-\rho^2}\left\{\rho + \frac{\rho}{1-\rho^2}(x^2-2\rho xy+y^2)+xy\right\}\\
&\geq xy\phi_2((x,y); (0,0), R),
\end{align*}
for $\rho\in (0,1)$.
Then, we have 
\begin{align*}
\frac{\partial}{\partial \rho}D_{\mu,\tau}(\rho)
\geq \iint xy\sigma(x)\sigma(y)\phi_2((x,y); (0,0), R)dxdy> 0,
\end{align*}
since $\rho_{ii'}>0$ and $x\sigma(x)$ is an increasing function. 
Hence, the derivative $\partial D_{\mu,\tau}(\rho)/\partial \rho$ is positive for $\rho\in (0,1)$.

\end{document}